\documentclass[11pt]{article}

\usepackage[final]{acl}

\usepackage{times}
\usepackage{latexsym}

\usepackage[T1]{fontenc}

\usepackage[utf8]{inputenc}

\usepackage{microtype}

\usepackage{inconsolata}

\usepackage{graphicx}

\usepackage{booktabs}
\usepackage{amsmath}
\usepackage{amssymb}
\usepackage{multirow}
\usepackage{array}
\usepackage{xspace}
\usepackage{enumitem}

\newcommand{\syn}{ReMixer\xspace}
\newcommand{\train}{Redapter\xspace}
\newcommand{\embed}{ReasonEmbed\xspace}
\newcommand{\bright}{BRIGHT\xspace}
\newcommand{\rrmed}{R2MED\xspace}

\title{ReasonEmbed: Enhanced Text Embeddings for Reasoning-Intensive Document Retrieval}

\author{
 \textbf{Jianlyu Chen\textsuperscript{1,2,4}}~~
 \textbf{Junwei Lan\textsuperscript{1,2,4}}~~
 \textbf{Chaofan Li\textsuperscript{2,3}}~~
 \textbf{Defu Lian\textsuperscript{1,4}$^{\ast}$}~~
 \textbf{Zheng Liu\textsuperscript{2,5}\thanks{Corresponding authors}}
\\
 \textsuperscript{1}University of Science and Technology of China
\\
 \textsuperscript{2}Beijing Academy of Artificial Intelligence \ \ \
\\
 \textsuperscript{3}Beijing University of Posts and Telecommunications \ \ \
\\
 \textsuperscript{4}State Key Laboratory of Cognitive Intelligence \ \ \
 \textsuperscript{5}Hong Kong Polytechnic University \\
{\tt chenjianlv@mail.ustc.edu.cn} \ \
{\tt liandefu@ustc.edu.cn} \ \
{\tt zhengliu1026@gmail.com}
\\
}

\begin{document}
\maketitle
\begin{abstract}
In this paper, we introduce \textbf{ReasonEmbed}, a novel text embedding model developed for reasoning-intensive document retrieval. Our work includes three key technical contributions. First, we propose \textbf{ReMixer}, a new data synthesis method that overcomes the triviality problem prevalent in previous synthetic datasets, enabling large-scale production of 82K high-quality training samples. Second, we design \textbf{Redapter}, a self-adaptive learning algorithm that dynamically adjusts training each sample's weight based on its reasoning intensity. This allows the model to effectively capture the complex semantic relationships between queries and documents. Third, we implement ReasonEmbed across multiple backbones of varying sizes, all of which achieve \textbf{superior performance} on reasoning-intensive retrieval tasks. Notably, our ReasonEmbed-Qwen3-8B model offers a record-high nDCG@10 score of 38.1 on the BRIGHT benchmark \cite{su2025bright}, which significantly outperforms existing text embedding models. We will fully open-source our created resources in ReasonEmbed to push forward the research advancement in this field\footnote{All resources will be available at \url{https://github.com/VectorSpaceLab/agentic-search/tree/main/ReasonEmbed}}. 
\end{abstract}

\section{Introduction}

With the rapid advancement of large language models (LLMs), autonomous AI agents have become increasingly popular across various real-world applications~\cite{yao2023react,wang2024survey,shinn2023reflexion}, such as personal assistants, software engineering, and scientific research. In many of these scenarios, AI agents require access to informative external references to ensure the generation of truthful answers. However, the complex semantic relationships that often exist between queries and documents in these emerging domains pose significant challenges for existing information retrieval systems. Recent studies~\cite{su2025bright} suggest that most traditional retrievers, like general-purpose text embeddings~\cite{neelakantan2022text,wang2024multilingual,xiao2024c,lee2024gecko} and BM25, struggle with these reasoning-intensive tasks, where effective retrieval often requires intensive reasoning operations. 

Despite the imperative demand, research on reasoning-intensive document retrieval encounters several fundamental challenges. A primary limitation lies in the scarcity of suitable training data. Most existing document retrieval datasets are curated from traditional applications~\cite{bajaj2016ms,kwiatkowski2019natural}, like web search and question answering, which differ substantially from the target problem in both query forms and domain knowledge. To alleviate data scarcity, recent studies have explored the use of synthetic datasets for developing retrieval systems~\cite{lee2024gecko,li2024making,wang2024e5mistral}. Building on this idea, preliminary efforts have focused on data synthesis strategies tailored to reasoning-intensive document retrieval. An early attempt was made by ReasonIR~\cite{shao2025reasonir}, where long-form queries and hard negatives are synthesized using scientific corpora. Subsequent research further advanced this direction by employing more sophisticated query generation methods or by mining potential queries from existing corpora~\cite{das2025rader,liu2025reasonrank,long2025diver}. However, current progress remains limited, as empirical evidence shows that these curated datasets yield only marginal gains over existing text embeddings. 

In this paper, we propose \textbf{\embed}, a new text embedding model for reasoning-intensive document retrieval based on innovations of how synthetic data is generated and used. Our work includes the following technical contributions. 

First, we design a novel data synthesis method, called \textbf{\syn}. Our study begins by identifying \textbf{triviality} as the key bottleneck in existing synthetic datasets. Specifically, synthetic data often exhibits overly direct relationships between queries and documents, where the relevance can be easily captured by surface patterns, like similar terms or overlapping keywords. To support this, Section~\ref{sec:data_triviality_evidence} presents evidence demonstrating that the triviality problem severely impairs the retrieval capability of fine-tuned embedding models. With this insight, we design a \textbf{three-stage workflow} comprising {conditioned query generation}, {source-excluded candidate mining}, and {reasoning-enhanced relevance annotation}. This effectively mitigates trivial cases while preserving the validity of the synthesized training data.

Second, we introduce a self-adaptive training method tailored for synthetic data, termed \textbf{\train}. Synthetic training samples exhibit varying levels of \textbf{reasoning intensity}, i.e., the degree of reasoning needed to capture the relationship between a query and its related document. Embedding models tend to reach performance saturation more quickly on samples with lower reasoning intensity. To address this, \train dynamically adjusts the weight of each training sample based on its estimated reasoning intensity, thereby enabling more effective utilization of the synthetic data. 

Third, we implement \embed based on multiple LLM backbones of varying model sizes, which achieve \textbf{state-of-the-art performance} on reasoning-intensive document retrieval tasks. Notably, our model built on Qwen3-4B reaches an nDCG@10 score of 37.1 on the \bright benchmark~\cite{su2025bright}, which already surpasses all existing text embeddings. While the Qwen3-8B based variant improves the performance to 38.1, yielding a significant improvement of almost +10 points over the cutting-edge baselines for this task. Extensive empirical analyses further validate the individual contributions of our synthesized data and self-adaptive training algorithm. 

To summarize, our research offers a preliminary yet insightful exploration of advancing IR techniques for newly emerged reasoning-intensive scenarios. Our results demonstrate that, despite unprecedented challenges, text embeddings continue to play a crucial role in such problems via effective optimization. The entire resources of this work, including the source code, curated dataset, and well-trained models, will be publicly released to facilitate future research in this field.

\section{Related Work}

In this section, the related works are reviewed from two aspects: text embeddings, and reasoning-intensive document retrieval.

\textbf{Text Embeddings}. Text embeddings have emerged as a major research direction and have been extensively studied in recent years. Early works~\cite{izacard2021unsupervised,wang2022text,li2023towards,xiao2024c,chen2024m3} focused on improving the capabilities of embedding models on general tasks through multi-stage training on large-scale datasets. More recent studies~\cite{ma2024repllama,wang2024e5mistral,li2024making,lee2025gemini,zhang2025qwen3embed} leverage powerful LLMs as backbone models for embeddings, achieving state-of-the-art performance on challenging benchmarks such as MTEB~\cite{muennighoff2022mteb} and MMTEB~\cite{enevoldsen2025mmteb}. To address the efficiency limitations of large LLM-based embeddings, following-up works further explore knowledge distillation from larger teacher models to enhance the performance of lightweight embeddings, thereby enabling more practical deployment in real-world applications~\cite{zhang2024jasper,askari2025hotelmatch,vera2025embeddinggemma}.

\textbf{Reasoning-Intensive Document Retrieval}. Reasoning-intensive document retrieval was first introduced in \bright~\cite{su2025bright}. Unlike traditional retrieval tasks, as represented by datasets such as MSMARCO~\cite{bajaj2016ms} and Natural Questions~\cite{kwiatkowski2019natural}, where keyword or semantic-based matching is often sufficient, reasoning-intensive document retrieval requires in-depth reasoning to identify relevant documents. This poses unprecedented challenges for existing general-purpose retrieval models. Recent research in this area can be broadly categorized into two directions. The first one aims to enhance first-stage retrieval performance on top of tailored embedding models and query rewriting methods~\cite{shao2025reasonir,das2025rader,long2025diver}, while the second one focuses on developing reasoning-enhanced re-ranking models tailored for reasoning-intensive tasks~\cite{niu2024judgerank,weller2025rank1,yang2025rank,zhuang2025rank,liu2025reasonrank,lan2025retro}. Despite the current progresses, there still remain fundamental challenges regarding the effectiveness of first-stage retrieval and the curation of training data.

\begin{figure*}[!t]
  \centering
  \includegraphics[width=1.0\linewidth]{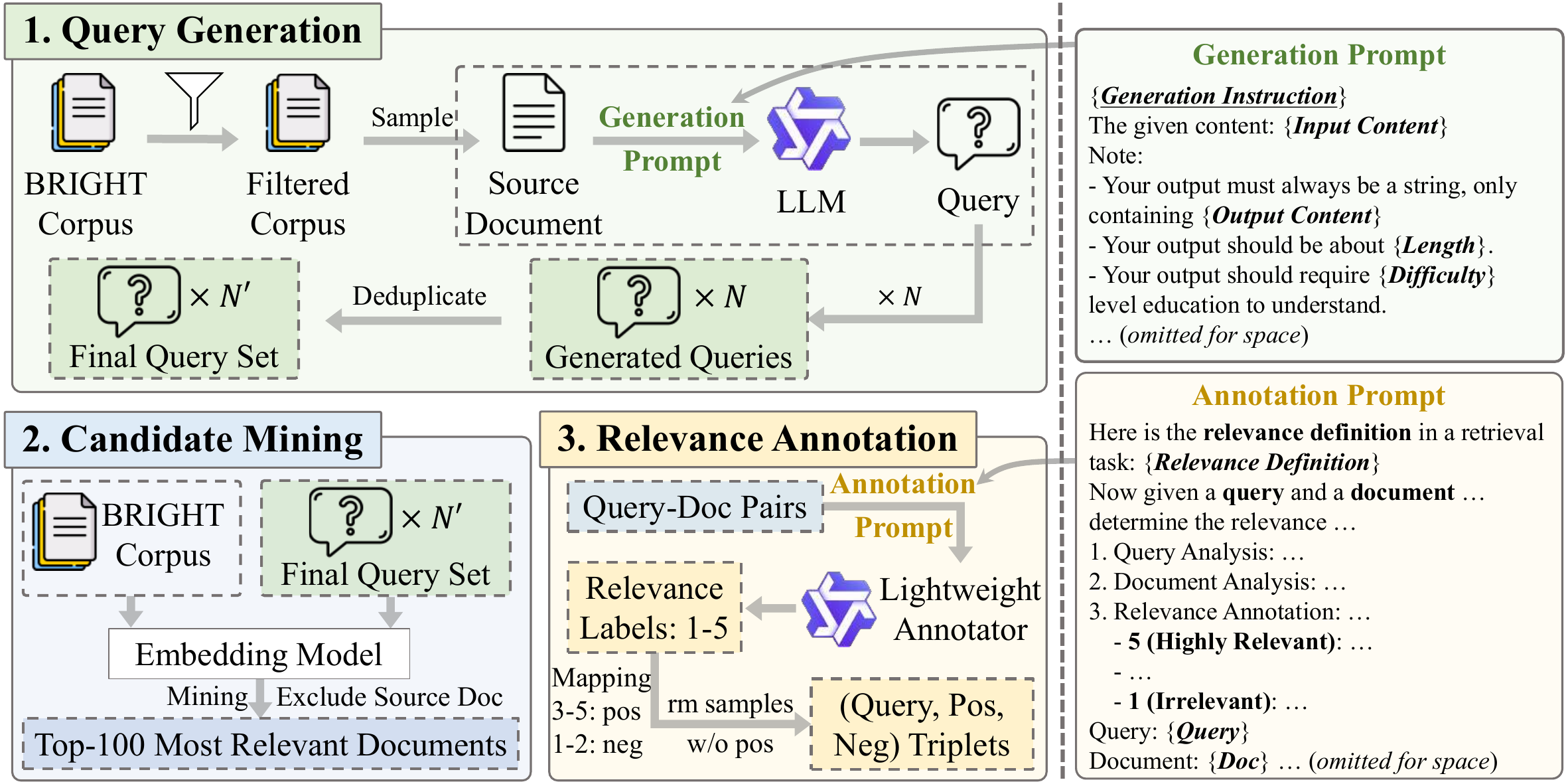}
  \caption{The three-stage data synthesis workflow of \syn. The full prompts used in the data synthesis process are available in Appendix~\ref{sec:appendix:data_synthesis}.}
  \label{fig:data_synthesis}
  \vspace{-5pt}
\end{figure*}

\section{Data Synthesis: \syn} 

To ensure optimal data usability, the synthetic data is expected to satisfy the following properties. 1) The generated queries should require reasoning to address their information needs. 2) The generated queries should exhibit sufficient diversity in terms of forms and domains. 3) The relationships between queries and their corresponding positive and hard-negative documents should be accurately labeled. Guided by these principles, the following data synthesis workflow is designed (Figure~\ref{fig:data_synthesis}). 

\subsection{Query Generation} 

We propose a conditioned query generation framework which prompts LLMs to produce quality and diversified queries. Our method begins by sourcing knowledge-rich corpora, which serve as a foundation for generating complex queries that call for intensive reasoning to address. In this work, we directly leverage the 12 datasets included in \bright~\cite{su2025bright}, which span diverse and closely related domains such as science, mathematics, and programming. We pre-process the sourced corpora by removing documents that are irrelevant to the labeled domain of their respective datasets. Furthermore, to prevent data leakage, we exclude any documents labeled as positives for the evaluation queries in the \bright benchmark. 

With the preprocessed datasets, we design a prompt template (shown in Figure~\ref{fig:data_synthesis}) to instruct the Qwen2.5-72B-Instruct~\cite{qwen2.5} model for query generation. The prompt incorporates three key design elements:
1) an explicit generation instruction, which ensures the production of reasoning-necessitated queries;
2) query length sampling, which promotes the creation of long-form queries with diverse lengths;
and 3) user education level sampling, which introduces variation in linguistic style and the depth of knowledge reflected in the generated queries. 

\subsection{Candidate Mining}\label{sec:doc_mining_annotation}

Although widely adopted in existing approaches, directly using source documents as positive samples often leads to trivial connections with the generated queries. To address this problem, we propose to exempt the source documents and instead introduce documents that are different in form but correlated in essence as positive candidates for each generated query. Specifically, for each query ($q$), we employ off-the-shelf retrievers ($\phi(\cdot)$) to mine candidate documents, formally defined as: $\mathcal{C}_q \leftarrow \text{Top-k}\{ \phi(q,d) \mid D / d^*_q \}$, where $D / d^*_q$ denotes the corpus excluding the source document $d^*_q$ associated with query $q$.  

\subsection{Relevance Annotation}

We further annotate the mined candidates, identifying positive documents for each query while treating the remaining ones as hard negatives. Given that the generated queries are designed to express complex information needs associated with knowledge-rich documents, the annotation process is conducted based on in-depth reasoning.

To achieve this, we perform \textit{reasoning-enhanced relevance annotation} with optimized treatments. On one hand, inspired by the successful practice in JudgeRank~\cite{niu2024judgerank}, the annotation workflow is formulated as a three-stage process: 1) \textit{Query analysis}, which examines the underlying information need expressed in the query; 2) \textit{Document analysis}, which assesses the detailed knowledge contained in the document; 3) \textit{Relevance annotation}, which determines the degree to which the document satisfies the query’s information need. The prompt template is shown in Figure~\ref{fig:data_synthesis}.

On the other hand, we leverage state-of-the-art reasoning LLMs as the backbone of the annotator. Considering that the annotation process consumes a huge amount of tokens, directly using a large-scale reasoning LLM is prohibitively expensive. To address this, we employ a tailored lightweight LLM via distillation for cost-effective processing. Specifically, a student annotator based on Qwen3-8B is fine-tuned using reasoning trajectories from Qwen3-235B-A22B-Instruct-2507~\cite{yang2025qwen3}. Detailed settings about these implementations are provided in Appendix~\ref{sec:appendix:data_synthesis}.

\subsection{Synthesization Result} 

As shown in Figure~\ref{fig:data_synthesis}, after the relevance annotation stage, each candidate document for one query has a relevance label in $\{1, 2, 3, 4, 5\}$. Documents with labels in $\{3, 4, 5\}$ are used as positives and those with labels in $\{1, 2\}$ are used as negatives. The queries without any positive document are filtered out of the final dataset. The statistics of the final synthetic dataset are summarized in Table~\ref{tab:data_stat}. In total, 95,960 raw queries were initially generated. After the annotation process, 81,659 valid queries remained, as those with no valid positive documents found in the sourced corpus were filtered out. In the final dataset, each query is associated with 12 positive documents on average, providing rich supervision signals for training retrieval models. The average query length reaches 221 tokens, which is substantially longer than those in traditional datasets and thus aligns well with the requirements of reasoning-intensive retrieval tasks. 

\begin{table}[!t]
    \centering
    \small
    
    \setlength{\tabcolsep}{3pt}
    \resizebox{0.48\textwidth}{!}{
        \begin{tabular}{l|c|cc|ccc}
        \toprule
        \multirow{2}{*}{\textbf{Source}} & \textbf{\#Query} & \multicolumn{2}{c|}{\textbf{Avg. \#Docs/Q}} & \multicolumn{3}{c}{\textbf{Avg. \#Tokens}} \\
        \cmidrule{3-7}
         & \textbf{(Final / Raw)} & \textbf{Pos} & \textbf{Neg} & \textbf{Q} & \textbf{Pos} & \textbf{Neg} \\
        \midrule
        \multicolumn{7}{l}{\textit{\textbf{StackExchange (7)}}} \\
        \midrule
        Bio.            & 7,470 / 9,180 & 8  & 90 & 118 & 159 & 111 \\
        Earth           & 8,492 / 9,917 & 8  & 90 & 112 & 489 & 178 \\
        Econ.           & 7,147 / 7,837 & 14 & 84 & 117 & 428 & 321 \\
        Psy.            & 5,412 / 5,987 & 10 & 88 & 119 & 471 & 311 \\
        Rob.            & 7,159 / 8,015 & 8  & 90 & 151 & 364 & 167 \\
        Stack.          & 6,170 / 6,873 & 12 & 86 & 178 & 864 & 699 \\
        Sus.            & 4,493 / 5,031 & 13 & 85 & 106 & 368 & 187 \\
        \midrule
        \multicolumn{7}{l}{\textit{\textbf{Coding (2)}}} \\
        \midrule
        Leet.           & 8,640 / 9,979 & 11 & 87 & 713 & 378 & 335 \\
        Pony            & 3,261 / 3,287 & 10 & 88 & 398 & 156 & 151 \\
        \midrule
        \multicolumn{7}{l}{\textit{\textbf{Math (3)}}} \\
        \midrule
        AoPS            & 7,370 / 9,939 & 22 & 76 & 159 & 305 & 329 \\
        TheoQ.          & 7,184 / 9,919 & 21 & 78 & 155 & 246 & 302 \\
        TheoT.          & 8,861 / 9,996 & 5  & 93 & 259 & 445 & 513 \\
        \midrule
        \textbf{Total}  & 81,659 / 95,960 & 12 & 86 & 221 & 383 & 308 \\
        \bottomrule
        \end{tabular}
    }
    \caption{Statistics of the synthetic dataset. \textbf{\#Query (Final / Raw)}: number of training queries (final / raw). \textbf{Avg. \#Docs/Q}: average number of positive or negative documents per query. \textbf{Avg. \#Tokens}: average number of tokens per data instance.}
    \label{tab:data_stat}
    \vspace{-5pt}
\end{table}

\section{Training Method: \train} 

With the synthetic dataset constructed, the embedding model is trained to discriminate positive documents from negative ones (including both in-batch negatives and the hard negatives provided in the synthetic dataset) for each query. The training objective follows the standard InfoNCE loss, which is formulated as:  
\begin{equation}\label{eq:info-nce}
    \min. \ \  \mathcal{L}_{q, D} = - \log \frac{\exp(\langle q,d^+ \rangle/\tau)}{\sum_{d'\in D}\exp(\langle q,d' \rangle /\tau)}, 
\end{equation}
where $D$ includes one positive $d^+$ and $|D|-1$ negatives, $\langle \cdot, \cdot \rangle$ denotes the dot-product similarity between the embeddings of a query $q$ and a document $d$, and $\tau$ is the temperature parameter.  
Since the embedding model must learn to capture the subtle semantic relationships between queries and documents, it is crucial to expose the model to sufficiently challenging training samples. Therefore, although reasoning-augmented queries, such as those produced by GPT-4 in \bright~\cite{su2025bright}, can improve retrieval performance at test time, we retain the original forms of the generated queries during training rather than rewriting them.  

\subsection{Reasoning Intensity} 

Although all training samples are produced through the same data synthesis pipeline, they exhibit varying levels of difficulty due to differences in their source documents and inherent randomness during the generation process. Easier samples typically involve relatively straightforward relationships between queries and documents, which can be captured without substantial reasoning and are learned quickly by the model. In contrast, harder samples embody more intricate semantic relationships that require deeper reasoning to resolve and therefore must be learned more patiently.

To capture this distinction, we define the concept of \textbf{reasoning intensity} to reflect such distinctions between training samples. Specifically, the hardness of a sample is characterized by the extent to which reasoning contributes to distinguishing relevant from irrelevant documents. For easy samples, the relevance can be determined through surface-level matching, making additional reasoning less beneficial. In contrast, hard samples depend heavily on multi-step reasoning, where additional reasoning operations significantly benefit the discrimination of the correct relationships. To quantify this property, we leverage the query-rewriting template from \bright~\cite{su2025bright} to generate a reasoning-augmented query $q'$ for each raw query $q$. The reasoning intensity of a sample $s = (q, D)$ is defined as the contrast between the query–doc similarity computed with and without reasoning:
\begin{equation}\label{eq:reasoning_intensity}
    \mathrm{RI}_\theta(s) = \min\left( \mathcal{L}_{q,D} / \mathcal{L}_{q',D}, \kappa \right),
\end{equation}
where $\theta$ is real-time parameter of the embedding model, $\mathcal{L}$ denotes the InfoNCE loss defined in Eq.~\ref{eq:info-nce}, and $\kappa$ is the hyperparameter used for truncating too large reasoning intensity scores. A larger $\mathrm{RI}_\theta(s)$ value indicates a stronger impact of reasoning, as the reasoning-augmented query ($q'$) substantially reduces the loss $\mathcal{L}_{q',D}$ compared to its non-reasoning counterpart $\mathcal{L}_{q,D}$. 

\begin{table*}[!ht]
    \centering
    
    \setlength{\tabcolsep}{3.5pt}
    \resizebox{1.0\textwidth}{!}{
        \begin{tabular}{l|c|c|ccccccc|cc|ccc}
            \toprule
            \multirow{2}{*}{\textbf{Models}} & \multirow{2}{*}{\textbf{Size}} & \multirow{2}{*}{\textbf{Avg.}} & \multicolumn{7}{c|}{\textbf{StackExchange}} & \multicolumn{2}{c|}{\textbf{Coding}} & \multicolumn{3}{c}{\textbf{Theorem-based}} \\
            \cmidrule{4-15}
            & & & \textbf{Bio.} & \textbf{Earth.} & \textbf{Econ.} & \textbf{Psy.} & \textbf{Rob.} & \textbf{Stack.} & \textbf{Sus.} & \textbf{Leet.} & \textbf{Pony} & \textbf{AoPS} & \textbf{TheoQ.} & \textbf{TheoT.} \\
            \midrule
            \multicolumn{15}{l}{\textbf{{General-purpose methods}}} \\
            \midrule
            BM25                        & -  & 14.5 & 18.9 & 27.2 & 14.9 & 12.5 & 13.6 & 18.4 & 15.0 & 24.4 & 7.9  & 6.2  & 10.4 & 4.9  \\
            OpenAI-3-Large              & -  & 17.9 & 23.3 & 26.7 & 19.5 & 27.6 & 12.8 & 14.3 & 20.5 & 23.6 & 2.4  & 8.5  & 23.5 & 11.7 \\
            Google-Gecko-1B-768         & 1B & 20.0 & 22.7 & 34.8 & 19.6 & 27.8 & 15.7 & 20.1 & 17.1 & 29.6 & 3.6  & 9.3  & 23.8 & 15.9 \\
            GritLM-7B                   & 7B & 21.0 & 24.8 & 32.3 & 18.9 & 19.8 & 17.1 & 13.6 & 17.8 & 29.9 & 22.0 & 8.8  & 25.2 & 21.2 \\
            gte-Qwen2-7B-instruct       & 7B & 23.5 & 34.1 & 42.6 & 18.2 & 27.4 & 13.2 & 17.3 & 20.9 & 30.4 & 2.2  & 13.3 & 30.6 & 32.6 \\
            Qwen3-Embedding-4B$^\ast$   & 4B & 21.8 & 17.8 & 34.7 & 16.9 & 23.3 & 12.5 & 16.2 & 16.8 & 35.7 & 1.4  & 9.8  & 35.5 & 41.5 \\
            Qwen3-Embedding-8B$^\ast$   & 8B & 22.8 & 21.0 & 33.0 & 18.4 & 26.1 & 15.7 & 19.4 & 17.3 & 33.8 & 1.2  & 9.4  & 39.2 & 39.3 \\
            Qwen3-4B-ms$^\dag$          & 4B & 19.4 & 16.8 & 34.9 & 16.0 & 19.4 & 20.1 & 18.8 & 11.1 & 33.3 & 9.1  & 8.7  & 27.0 & 18.0 \\
            Qwen3-8B-ms$^\dag$          & 8B & 18.7 & 15.9 & 35.6 & 16.1 & 19.1 & 20.5 & 19.2 & 11.7 & 29.2 & 9.5  & 8.4  & 22.2 & 17.0 \\
            Llama-3.1-8B-ms$^\dag$      & 8B & 16.1 & 12.2 & 29.3 & 14.3 & 16.7 & 14.7 & 14.4 & 11.6 & 28.9 & 3.5  & 8.7  & 23.7 & 14.9 \\
            \midrule
            \multicolumn{15}{l}{\textbf{{Tailored methods for reasoning-intensive retrieval}}} \\
            \midrule
            ReasonIR-8B                 & 8B & 24.4 & 26.2 & 31.4 & 23.3 & 30.0 & 18.0 & 23.9 & 20.5 & 35.0 & 10.5 & 14.7 & 31.9 & 27.2 \\
            RaDeR-gte-Qwen2-7B          & 7B & 25.5 & 34.6 & 38.9 & 22.1 & 33.0 & 14.8 & 22.5 & 23.7 & 37.3 & 5.0  & 10.2 & 28.4 & 35.1 \\
            Seed-1.5-Embedding          & -  & 27.2 & 34.8 & 46.9 & 23.4 & 31.6 & 19.1 & 25.4 & 21.0 & 43.2 & 4.9  & 12.2 & 33.3 & 30.5 \\
            DIVER-Retriever             & 4B & 28.9 & 41.8 & 43.7 & 21.7 & 35.3 & 21.0 & 21.2 & 25.1 & 37.6 & 13.2 & 10.7 & 38.4 & 37.3 \\
            \midrule
            \multicolumn{15}{l}{\textbf{{\embed from basic contrastive learning (using InfoNCE loss)}}} \\
            \midrule
            \embed-Qwen3-4B             & 4B & 35.3 & 51.8 & 53.3 & 34.1 & 42.6 & 31.1 & 32.1 & 35.6 & 32.7 & 11.6 & 13.0 & 40.8 & 45.2 \\
            \embed-Qwen3-8B             & 8B & 37.1 & 54.4 & 55.4 & 33.8 & 45.2 & 32.0 & 34.3 & 37.3 & 32.3 & 18.7 & 13.3 & 41.2 & 47.6 \\
            \embed-Llama-3.1-8B         & 8B & 34.9 & 55.5 & 53.8 & 36.4 & 45.6 & 29.8 & 35.1 & 38.6 & 29.4 & 12.5 & 9.5  & 36.7 & 35.7 \\
            \midrule
            \multicolumn{15}{l}{\textbf{{\embed from \train (using the self-adaptive RI-InfoNCE loss)}}} \\
            \midrule
            \embed-Qwen3-4B             & 4B & 37.1 & 55.4 & 54.5 & 34.9 & 46.9 & 34.0 & 36.1 & 37.4 & 34.5 & 13.6 & 11.3 & 41.4 & 45.1 \\
            \embed-Qwen3-8B             & 8B & 38.1 & 55.5 & 56.6 & 36.2 & 47.4 & 35.3 & 36.6 & 39.1 & 33.6 & 16.4 & 12.5 & 41.4 & 47.2 \\
            \embed-Llama-3.1-8B         & 8B & 36.2 & 55.4 & 56.2 & 35.2 & 48.5 & 32.1 & 37.3 & 41.1 & 28.8 & 16.8 & 9.1  & 37.9 & 36.6 \\
            \bottomrule
        \end{tabular}
    }
    \caption{Main evaluation results (nDCG@10) on the \bright benchmark (using original queries). For the baselines,``$\ast$'' denotes well-trained models released by existing studies and evaluated by us; ``$\dag$'' indicates reproduced methods in this work. Other results are directly taken from the literature~\cite{su2025bright,das2025rader,shao2025reasonir,long2025diver}.}
    \label{tab:main_results_bright}
    \vspace{-5pt}
\end{table*}

\subsection{Self-Adaptive Learning}

With the introduction of reasoning intensity, we introduce a new self-adaptive learning method which switches to minimize the \textbf{RI-InfoNCE} loss defined as the following equation: 
\begin{equation}\label{eq:ri-info-nce}
    \mathcal{L}_{\mathrm{RI}} = \sum_{s=(q,D), s\in B}f(\mathrm{RI}_{\theta}(s),B)*\mathcal{L}_{q,D},
\end{equation}
where $s$ is a training sample within a batch $B$, and $f(\cdot)$ is a normalization function that scales the reasoning-intensity scores within the batch: 
\begin{equation}
    f(\mathrm{RI}_{\theta}(s),B) = \mathrm{RI}_{\theta}(s) / \sum\nolimits_{s' \in B}\mathrm{RI}_{\theta}(s'). 
\end{equation}

The embedding model is initialized from a checkpoint pretrained on the MSMARCO dataset~\cite{bajaj2016ms}. It is then optimized based on the synthetic dataset, with RI-InfoNCE adopted as the training objective. During the training process, samples with higher reasoning intensity receive greater weights, allowing the model to allocate more learning capacity to reasoning-intensive cases while retaining moderate exposure to simpler examples. This self-adaptive weighting strategy facilitates a more efficient and balanced learning process that enhances the model’s representation capability in complex retrieval scenarios. Notably, \train does not increase the training computation cost significantly. We provide the corresponding evidence in Appendix~\ref{sec:appendix:training_cost_analysis}.

\section{Experiments}

In this section, we conduct experiments for the following research questions:

\noindent
\textbf{RQ1}: How effective is \embed in reasoning-intensive document retrieval compared with existing text embedding models?

\noindent
\textbf{RQ2}: What is the impact of the synthetic dataset on \embed's overall performance?

\noindent
\textbf{RQ3}: How does the proposed training method contribute to \embed's overall performance? 

\noindent
\textbf{RQ4}: How consistent and accurate are the annotation labels generated by the distilled annotator?

\noindent
\textbf{RQ5}: What insights can be drawn from the analysis of important implementation details?

\begin{table*}[!ht]
    \centering
    \small
    
    \setlength{\tabcolsep}{4.5pt}
    \resizebox{1.0\textwidth}{!}{
        \begin{tabular}{l|c|c|ccc|ccc|cc}
            \toprule
            \multirow{2}{*}{\textbf{Models}} & \multirow{2}{*}{\textbf{Size}} & \multirow{2}{*}{\textbf{Avg.}} & \multicolumn{3}{c|}{\textbf{Q\&A Reference}} & \multicolumn{3}{c|}{\textbf{Clinical Evidence}} & \multicolumn{2}{c}{\textbf{Clinical Case}} \\
            \cmidrule{4-11}
            & & & \textbf{Biology} & \textbf{Bioin.} & \textbf{MedS.} & \textbf{MedE.} & \textbf{MedD.} & \textbf{PMCT.} & \textbf{PMCC.} & \textbf{IIYiC.} \\
            \midrule
            \multicolumn{11}{l}{\textbf{\textit{General-purpose methods}}} \\
            \midrule
            BM25                        & -  & 15.13 & 19.19 & 21.55 & 19.68 & 0.66  & 2.55  & 23.69 & 21.66 & 12.02 \\
            OpenAI-3-Large              & -  & 28.57 & 23.82 & 40.51 & 44.05 & 11.78 & 15.01 & 47.43 & 28.87 & 17.12 \\
            GritLM-7B                   & 7B & 31.12 & 24.99 & 43.98 & 45.94 & 12.32 & 19.86 & 39.88 & 37.08 & 24.94 \\
            NV-Embed-v2                 & 7B & 31.43 & 27.15 & 50.10 & 47.81 & 10.90 & 16.72 & 44.05 & 39.91 & 14.81 \\
            gte-Qwen2-7B-instruct$^\ast$& 7B & 32.56 & 33.18 & 45.53 & 49.91 & 13.41 & 17.10 & 48.19 & 32.13 & 21.07 \\
            Qwen3-Embedding-4B$^\ast$   & 4B & 31.75 & 18.16 & 47.73 & 43.37 & 17.25 & 22.90 & 47.19 & 33.65 & 23.76 \\
            Qwen3-Embedding-8B$^\ast$   & 8B & 34.22 & 21.37 & 50.15 & 46.65 & 18.91 & 27.06 & 48.76 & 37.29 & 23.60 \\
            Qwen3-4B-ms$^\dag$          & 4B & 23.09 & 16.05 & 37.30 & 33.67 & 5.28  & 8.79  & 31.55 & 29.63 & 22.43 \\
            Qwen3-8B-ms$^\dag$          & 8B & 24.15 & 17.02 & 38.11 & 39.33 & 5.37  & 9.51  & 31.85 & 30.88 & 21.11 \\
            Llama-3.1-8B-ms$^\dag$      & 8B & 22.34 & 13.17 & 34.76 & 34.30 & 5.64  & 8.83  & 33.10 & 30.68 & 18.23 \\
            \midrule
            \multicolumn{11}{l}{\textbf{{Tailored methods for reasoning-intensive retrieval}}} \\
            \midrule
            ReasonIR-8B$^\ast$          & 8B & 27.94 & 26.16 & 44.84 & 39.28 & 11.21 & 14.92 & 36.56 & 29.20 & 21.37 \\
            RaDeR-gte-Qwen2-7B$^\ast$   & 7B & 35.19 & 36.02 & 53.58 & 50.32 & 15.41 & 20.31 & 49.94 & 34.31 & 21.83 \\
            DIVER-Retriever$^\ast$      & 4B & 32.23 & 39.27 & 51.68 & 50.81 & 13.94 & 16.10 & 38.56 & 24.69 & 22.81 \\
            \midrule
            \multicolumn{11}{l}{\textbf{{\embed from basic contrastive learning (using InfoNCE loss)}}} \\
            \midrule
            \embed-Qwen3-4B             & 4B & 39.94 & 49.50 & 62.59 & 60.26 & 20.43 & 24.75 & 49.27 & 30.40 & 22.32 \\
            \embed-Qwen3-8B             & 8B & 42.36 & 51.65 & 65.17 & 65.89 & 20.67 & 27.98 & 51.71 & 33.79 & 22.01 \\
            \embed-Llama-3.1-8B         & 8B & 41.98 & 53.04 & 63.24 & 63.89 & 22.20 & 29.43 & 51.90 & 32.90 & 19.20 \\
            \midrule
            \multicolumn{11}{l}{\textbf{{\embed from \train (using the self-adaptive RI-InfoNCE loss)}}} \\
            \midrule
            \embed-Qwen3-4B             & 4B & 41.16 & 52.45 & 64.28 & 62.58 & 20.83 & 26.21 & 48.03 & 32.50 & 22.38 \\
            \embed-Qwen3-8B             & 8B & 43.18 & 54.01 & 66.33 & 67.64 & 20.93 & 27.96 & 51.38 & 33.76 & 23.43 \\
            \embed-Llama-3.1-8B         & 8B & 42.76 & 53.52 & 64.37 & 63.82 & 20.40 & 29.67 & 51.86 & 34.55 & 23.88 \\
            \bottomrule
        \end{tabular}
    }
    \caption{Main evaluation results (nDCG@10) on the \rrmed benchmark (using original queries). For the baselines, ``$\ast$'' denotes well-trained models released by existing studies and evaluated by us; ``$\dag$'' indicates reproduced methods in this work. Other results are directly taken from \citet{li2025r2med}.}
    \label{tab:main_results_r2med}
    \vspace{-5pt}
\end{table*}

\subsection{Setup}\label{sec:setup} 

The basic settings of the experimental studies are presented as follows. 

\textbf{Backbones}. We adopt Qwen3-8B~\cite{yang2025qwen3} as the default backbone for all experiments. Besides, we use Qwen3-4B and Llama-3.1-8B~\cite{grattafiori2024llama} to evaluate \embed's effectiveness across other popular models. 

\textbf{Training}. The entire 82K synthesized training samples are used for fine-tuning the backbone models. The training is conducted for one single epoch, where the RI-InfoNCE loss function is applied after a warm-up stage of 100 training steps with the basic InfoNCE loss. Detailed training parameters are provided in Appendix~\ref{sec:appendix:training_parameter}. 

\textbf{Evaluation}. We use \bright~\cite{su2025bright}, which includes 12 datasets spanning three domains, as the main evaluation benchmark. To further assess \embed’s out-of-domain performance, we also incorporate \rrmed~\cite{li2025r2med}, a \bright-like reasoning-intensive retrieval benchmark specifically curated for healthcare scenarios. We also demonstrate the effectiveness of our synthetic data on non-reasoning-intensive retrieval tasks in Appendix~\ref{appendix:non_reasoning_intensive_performance}.

\textbf{Baselines}. We compare against two categories of baselines: 1) general-purpose baseline retrievers and 2) tailored methods for reasoning-intensive retrieval. The experiment results for \bright and \rrmed are presented in Table~\ref{tab:main_results_bright} and Table~\ref{tab:main_results_r2med}, respectively. Details on the baseline methods are provided in Appendix~\ref{appendix:experiment:baselines}. 

\subsection{Main Results (RQ1-RQ3)}\label{sec:main_results}  

The experimental results on the \bright benchmark are presented in Table~\ref{tab:main_results_bright}, where all variants of \embed demonstrate superior performance. In particular, \embed-Qwen3-4B achieves an average nDCG@10 score of 37.1, surpassing all existing baselines. \embed-Qwen3-8B further improves performance to 38.1, establishing a substantial margin over previous methods. Complementary results on the \rrmed benchmark, shown in Table~\ref{tab:main_results_r2med}, reveal that \embed maintains a strong advantage over all baselines. Notably, \embed-Qwen3-8B achieves an nDCG@10 score of 43.18, which significantly improves upon the MSMARCO-finetuned baseline Qwen3-8B-ms (24.15) and outperforms the competitive baseline, DIVER-Retriever (32.23), by more than 10 points. It's worth noting that \rrmed consists of healthcare-related queries and documents, which is substantially different from the data used to develop \embed. Thus, it highlights the model’s strong out-of-domain generalizability. 
Overall, the above results provide compelling evidence for the effectiveness of \embed. 

In addition, the above results also provide crucial insights to the effectiveness of both data synthesis strategy and self-adaptive training method. 

First, the synthetic dataset generated by \syn plays a critical role in establishing \embed’s superior performance. Specifically, models fine-tuned on the synthetic data, including \embed-Qwen3-4B, \embed-Qwen3-8B, and \embed-Llama-3.1-8B, achieve significant improvements over their MSMARCO-finetuned counterparts (Qwen3-4B-ms, Qwen3-8B-ms, and Llama-3.1-8B-ms). Moreover, these models also outperform general-purpose embedding baselines built on the same backbone encoders, such as Qwen3-Embedding-4B and Qwen3-Embedding-8B, both of which have been extensively trained on existing text retrieval datasets. This overwhelming advantage clearly demonstrates the effectiveness of the synthetic dataset curated by \syn. 

Second, the self-adaptive training algorithm \train further enhances the overall performance of \embed. Notably, the \train-based variants consistently outperform their counterparts trained with conventional contrastive learning. These results validate the effectiveness of our self-adaptive training approach and underscore the crucial role of reasoning-intensive samples, a key factor emphasized throughout both the data synthesization and training stages. 

\subsection{Annotation Quality Assessment (RQ4)}

To assess the annotation quality of the distilled annotator, we evaluate its consistency with the teacher model and compute its accuracy based on human judgments from three authors of this paper.

\begin{table}[!t]
    \centering
    \small
    
    \setlength{\tabcolsep}{15pt}
        \begin{tabular}{l|c}
            \toprule
            \textbf{Model} & \textbf{Cohen's Kappa} \\
            \midrule
            Distilled Annotator     & 0.6001  \\
            Zero-shot Base Model    & 0.4314  \\
            \bottomrule
        \end{tabular}
    \caption{Annotation consistency evaluation of the distilled annotator and the zero-shot base model against the teacher model.}
    \label{tab:ablation:annotation_consistency}
    \vspace{-5pt}
\end{table}

\textbf{Annotation Consistency}. We construct a held-out evaluation set by sampling 500 queries per task and 10 documents per query, yielding 60,000 test samples for annotation consistency evaluation. We first use the teacher model and the distilled annotator to annotate these test samples independently, and then evaluate the annotation consistency using Cohen's Kappa. For comparison, we also evaluate the agreement between the base model (Qwen3-8B) and the teacher model. As shown in Table~\ref{tab:ablation:annotation_consistency}, the distilled annotator obtains a Cohen's Kappa score of 0.6001 on average, showcasing substantial agreement with the teacher model. Moreover, in comparison with the base model's score of 0.4314, the distilled annotator achieves an absolute gain of 0.1687, demonstrating the effectiveness of the distillation process.

\begin{table}[!t]
    \centering
    \small
    
    \setlength{\tabcolsep}{16pt}
        \begin{tabular}{l|c}
            \toprule
            \textbf{Model}      & \textbf{Accuracy} \\
            \midrule
            Teacher Model       & 86.1\% (310/360) \\
            Distilled Annotator & 85.0\% (306/360) \\
            \bottomrule
        \end{tabular}
    \caption{Annotation accuracy evaluation of the distilled annotator and the teacher model.}
    \label{tab:ablation:annotation_accuracy}
    \vspace{-5pt}
\end{table}

\textbf{Annotation Accuracy}. We construct a held-out evaluation set by sampling 10 queries per task and 3 documents per query, yielding 360 test samples for annotation accuracy evaluation. We employ the distilled annotator and the teacher model to annotate these test samples independently. Then, human annotators validate the accuracy under a strict standard, where the label is considered accurate if and only if it aligns with human judgment. As shown in Table~\ref{tab:ablation:annotation_accuracy}, both the distilled annotator and the teacher model achieve high average accuracies (85.0\% and 86.1\%, respectively), which demonstrates the high quality of our final annotated training data.

\subsection{Ablation Study (RQ5)}

We conduct ablation study to analyze the detailed impact of key designs across different components of our framework, including 1) candidate mining and 2) relevance annotation in data synthesis, 3) reasoning-intensity computation methods in the self-adaptive training, and 4) the influence of scaling-up of the synthesized data. To eliminate the effect of the self-adaptive training algorithm, the basic InfoNCE loss is used for all ablation studies except that on reasoning-intensity computation methods.

\subsubsection{Candidate Mining}
\label{sec:ablation:data_synthesis}

We first examine the impact of the candidate mining method. As discussed in Section~\ref{sec:doc_mining_annotation}, our default implementation (denoted as \underline{\textsc{Default}}) mines candidate documents from the corpus while explicitly excluding the source documents to avoid trivial matches. 
In the ablation study, we introduce three alternative settings to analyze its effect: 
1) \underline{\textsc{Non-Anno}}, which replaces positive samples with their corresponding source documents and directly uses the mined negatives without additional annotation, 
essentially following the common practice adopted in prior studies~\cite{moreira2024nv}.
2) \underline{\textsc{Source}}, which substitutes the positive samples with the source documents while keeping the same set of negatives as in \textsc{Default}. 3) \underline{\textsc{Source-Pro}}, which further annotates the source documents and only kept those annotated as positive on the basis of \textsc{Source}.

The experiment results are presented in Table~\ref{tab:ablation:data_analysis}, where the following observations are made. First, when comparing \textsc{Default} with \textsc{Non-Anno}, \textsc{Source}, and \textsc{Source-Pro}, the alternative methods only achieve nDCG@10 scores of 16.1, 14.5, and 22.1, respectively, which are substantially lower than the performance of the default method. This clearly validates the necessity of excluding source documents when curating synthetic training samples. Second, comparing \textsc{Non-Anno}, \textsc{Source}, and \textsc{Source-Pro} with one another, we observe that the annotation process lifts the overall performance to 22.1 (\textsc{Source-Pro}), despite the triviality introduced by the source documents. This further highlights the critical role of relevance annotation in the data synthesis workflow.

\begin{table}[!t]
    \centering
    
    \setlength{\tabcolsep}{3pt}
    \resizebox{0.48\textwidth}{!}{
        \begin{tabular}{l|l|l|c}
            \toprule
            \textbf{Method} & \textbf{Positive} & \textbf{Negative} & \textbf{Avg.} \\
            \midrule
            \underline{\textsc{Default}}        & mined \& annotat.     & mined \& annotat. & 37.1 \\
            \underline{\textsc{Non-Anno}}       & source                & mined             & 16.1 \\
            \underline{\textsc{Source}}         & source                & mined \& annotat. & 14.5 \\
            \underline{\textsc{Source-Pro}}     & source \& annotat.    & mined \& annotat. & 22.1 \\
            \bottomrule
        \end{tabular}
    }
    \caption{Impact of candidate mining methods (evaluated based on the \bright benchmark).} 
    \label{tab:ablation:data_analysis}
    \vspace{-5pt}
\end{table}

\begin{table}[!t]
    \centering
    
    \setlength{\tabcolsep}{3pt}
    \resizebox{0.48\textwidth}{!}{
        \begin{tabular}{l|ccc|c}
            \toprule
            \textbf{Method} & \textbf{ROUGE-1} & \textbf{ROUGE-2} & \textbf{ROUGE-L} & \textbf{BRIGHT Score} \\
            \midrule
            \underline{\textsc{Default}}    & 0.2411 & 0.0536 & 0.1439 & 37.1 \\
            \underline{\textsc{Source}}     & 0.3216 & 0.1351 & 0.2167 & 14.5 \\
            \bottomrule
        \end{tabular}
    }
    \caption{ROUGE scores and BRIGHT performance comparison between the baseline source-as-positive method and our ReMixer method.} 
    \label{tab:ablation:data_triviality_evidence}
    \vspace{-5pt}
\end{table}

\label{sec:data_triviality_evidence}
Moreover, to illustrate the data triviality problem of the source-as-positive synthesis method, we compute and compare the ROUGE scores of two data sets: 1) the raw 95,960 generated query-positive pairs, where the positive is the source document used to generate the query, and 2) the final 81,659 synthesized query-positive pairs, where the positive is randomly sampled from all positives for that query. As presented in Table~\ref{tab:ablation:data_triviality_evidence}, \textsc{Default} yields consistently lower ROUGE scores compared to \textsc{Source}. Furthermore, the BRIGHT performance results indicate that the triviality problem severely impairs the retrieval capability of fine-tuned embedding models.

\subsubsection{Relevance Annotation}
\label{sec:ablation:relevance_annotation}

We next analyze the impact of data annotation by comparing different annotation methods. Our default method, which employs a distilled lightweight reasoning-LLM, is denoted as \underline{\textsc{Default}}. To assess its effectiveness, we introduce two alternative baselines. 1) \underline{\textsc{Zero-shot}}, which directly uses the original lightweight LLM without distillation from the teacher model; and 2) \underline{\textsc{Non-Reason}}, which disables the reasoning process and trains the lightweight LLM to output annotation scores directly (refer to Appendix~\ref{appendix:experiment:detailed_ablation_results} for details). 

\begin{table}[!t]
    \centering
    \small
    
    \setlength{\tabcolsep}{4pt}
        \begin{tabular}{l|l|l|c}
            \toprule
            \textbf{Method} & \textbf{Training} & \textbf{Reasoning} & \textbf{Avg.} \\
            \midrule
            \underline{\textsc{Default}}        & distilled       & w/ reasoning  & 37.1 \\
            \underline{\textsc{Zero-shot}}      & zero-shot       & w/ reasoning  & 32.4 \\
            \underline{\textsc{Non-Reason}}     & distilled       & w/o reasoning & 35.0 \\
            \bottomrule
        \end{tabular}
    \caption{Impact of on relevance annotation methods (evaluated based on the \bright benchmark).}
    \label{tab:ablation:relevance_annotation}
    \vspace{-5pt}
\end{table}

\begin{table}[!t]
    \centering
    \small
    
    \setlength{\tabcolsep}{20pt}
        \begin{tabular}{l|c}
            \toprule
            \textbf{LLM Reasoner} & \textbf{Avg.} \\
            \midrule
            GPT-4.1-mini (default)  & 38.1 \\
            Qwen3-32B               & 37.8 \\
            Qwen3-8B                & 37.5 \\
            Qwen3-4B                & 36.5 \\
            \bottomrule
        \end{tabular}
    \caption{Impact of reasoning intensity scores computed by different LLM reasoners (evaluated based on the \bright benchmark).}
    \label{tab:ablation:training}
    \vspace{-5pt}
\end{table}

The experiment results are reported in Table~\ref{tab:ablation:relevance_annotation}, from which the following observations are made. First, with the adoption of the distilled annotator, \textsc{Default} improves overall performance from 32.4 (\textsc{Zero-shot}) to 37.1. This suggests that lightweight LLMs lack sufficient reasoning capability for reasoning-enhanced relevance annotation, while distillation from a powerful teacher model effectively alleviates this limitation. Second, incorporating explicit reasoning further enhances performance, with \textsc{Default} improving the performance from 35.0 (\textsc{Non-Reason}) to 37.1. This underscores the importance of reasoning in accurately discriminating nuanced query–document relationships during data annotation.

\subsubsection{Reasoning-Intensity}
\label{sec:ablation:training}

We further investigate the impact of reasoning-intensity scores computed using different LLM reasoners. In this study, we incorporate four popular alternatives: GPT-4.1-mini (default), Qwen3-32B, Qwen3-8B, and Qwen3-4B. It is worth noting that the reasoning outputs, i.e., the reasoning-augmented queries, are generated in the offline stage, thus will not affect the efficiency of online training.
The experimental results are presented in Table~\ref{tab:ablation:training}. As observed, the default setting using GPT-4.1-mini achieves the best performance, whereas models relying on lighter LLMs such as Qwen3-8B and Qwen3-4B perform noticeably worse. This highlights that generating high-quality reasoning augmentations is critical for accurately estimating reasoning intensity, and that using sufficiently powerful reasoning LLMs is essential for ensuring its effectiveness. 

\subsubsection{Data Scaling}
\label{sec:ablation:data_size_scaling}

\begin{figure}[!t]
  \centering
  \includegraphics[width=1.0\linewidth]{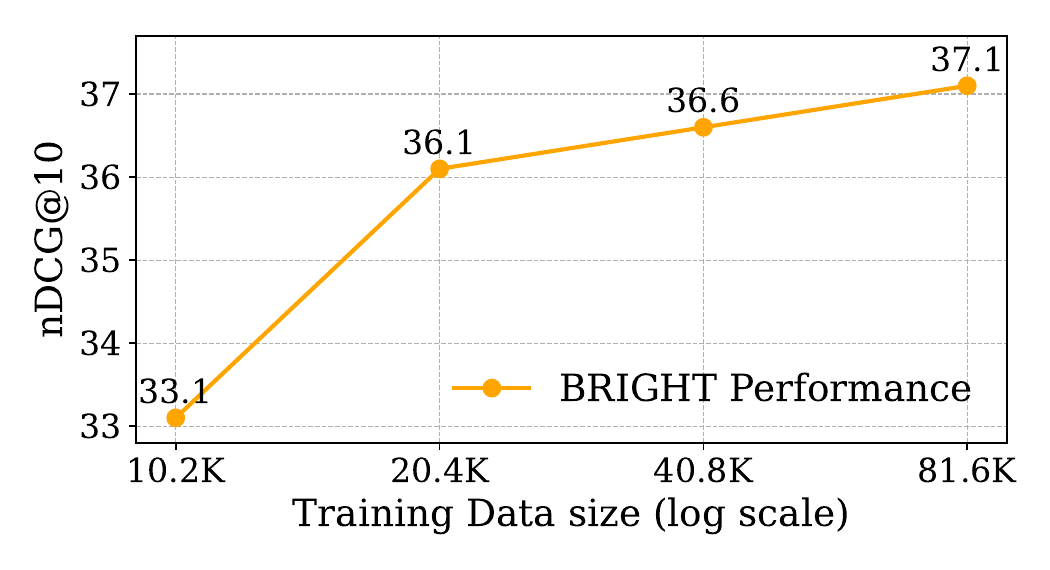}
  \caption{Impact of synthetic data size on retrieval accuracy (using basic contrastive learning for simplicity).}
  \label{fig:ablation:training_data_size_scale} 
  \vspace{-5pt}
\end{figure}

We finally investigate the impact of synthetic data size. Since our data synthesis process is fully automated, the dataset can be continuously expanded at negligible cost. In our experiments, \embed is fine-tuned on gradually enlarged datasets. As shown in Figure~\ref{fig:ablation:training_data_size_scale}, the retrieval performance improves substantially as the dataset grows from 10.2K to 81.6K samples. These results further demonstrate the value of our synthesized data and highlight the importance of automated data synthesis for scaling reasoning-intensive retrieval models. 

\section{Conclusion}

In this work, we present \embed, a novel embedding model designed for the emerging task of reasoning-intensive document retrieval. We first introduce \syn, a three-stage data synthesis workflow that generates diverse and reasoning-intensive training samples. Building on this, we further propose \train, a self-adaptive training method that dynamically weights samples based on their reasoning intensity, enabling more effective learning from the synthetic data. Finally, our extensive experiments on the \bright benchmark demonstrate that \embed achieves significant improvements over existing methods, while all introduced technical component substantially contribute to its superior performance.

\section*{Limitations}

While \embed demonstrates strong performance in reasoning-intensive document retrieval, there remain several problems to improve. First, although the data synthesis method significantly enhances retrieval accuracy, it is still constrained by the reasoning capacity of the underlying LLMs. This limitation could be mitigated in the future with more powerful LLMs, or with alternative approaches, such as crafted multi-agent workflows. Second, the current work primarily focuses on reasoning-intensive retrieval. Expanding training to include both reasoning-intensive and general retrieval data would make \embed more broadly applicable in practice. Third, our study is currently limited to domains covered by existing benchmarks, particularly \bright. Scaling to more diverse domains will further strengthen the model’s generalizability. Addressing these challenges will be a focus of our future research. 

\section*{Ethics Consideration}

Since our synthetic dataset is generated by LLM, it may inherit potential biases, toxicity, and other issues present in the LLM used during the generation process. Moreover, the corpora used in the generation process are derived from the real-world sources, they may contain sensitive content.

\section*{Acknowledgments}

This work is supported by grants from the National Natural Science Foundation of China (No. U24A20253).

\bibliography{custom}

\begin{thebibliography}{49}
\providecommand{\natexlab}[1]{#1}

\bibitem[{Askari et~al.(2025)Askari, Stergiadis, Gusev, and Beladev}]{askari2025hotelmatch}
Arian Askari, Emmanouil Stergiadis, Ilya Gusev, and Moran Beladev. 2025.
\newblock \href {https://doi.org/10.18653/v1/2025.acl-long.30} {{H}otel{M}atch-{LLM}: Joint multi-task training of small and large language models for efficient multimodal hotel retrieval}.
\newblock In \emph{Proceedings of the 63rd Annual Meeting of the Association for Computational Linguistics (Volume 1: Long Papers)}, pages 607--619, Vienna, Austria. Association for Computational Linguistics.

\bibitem[{Bajaj et~al.(2016)Bajaj, Campos, Craswell, Deng, Gao, Liu, Majumder, McNamara, Mitra, Nguyen et~al.}]{bajaj2016ms}
Payal Bajaj, Daniel Campos, Nick Craswell, Li~Deng, Jianfeng Gao, Xiaodong Liu, Rangan Majumder, Andrew McNamara, Bhaskar Mitra, Tri Nguyen, and 1 others. 2016.
\newblock Ms marco: A human generated machine reading comprehension dataset.
\newblock \emph{arXiv preprint arXiv:1611.09268}.

\bibitem[{Capari et~al.(2024)Capari, Azarbonyad, Tsatsaronis, Afzal, Dunham, and Kamps}]{capari2024knowledge}
Artemis Capari, Hosein Azarbonyad, Georgios Tsatsaronis, Zubair Afzal, Judson Dunham, and Jaap Kamps. 2024.
\newblock Knowledge acquisition passage retrieval: corpus, ranking models, and evaluation resources.
\newblock In \emph{International Conference of the Cross-Language Evaluation Forum for European Languages}, pages 74--87. Springer.

\bibitem[{Chen et~al.(2024)Chen, Xiao, Zhang, Luo, Lian, and Liu}]{chen2024m3}
Jianlyu Chen, Shitao Xiao, Peitian Zhang, Kun Luo, Defu Lian, and Zheng Liu. 2024.
\newblock \href {https://doi.org/10.18653/v1/2024.findings-acl.137} {{M}3-embedding: Multi-linguality, multi-functionality, multi-granularity text embeddings through self-knowledge distillation}.
\newblock In \emph{Findings of the Association for Computational Linguistics: ACL 2024}, pages 2318--2335, Bangkok, Thailand. Association for Computational Linguistics.

\bibitem[{Das et~al.(2025)Das, Nuallain, and Rahimi}]{das2025rader}
Debrup Das, Sam~O' Nuallain, and Razieh Rahimi. 2025.
\newblock Rader: Reasoning-aware dense retrieval models.
\newblock \emph{arXiv preprint arXiv:2505.18405}.

\bibitem[{Douze et~al.(2024)Douze, Guzhva, Deng, Johnson, Szilvasy, Mazar{\'e}, Lomeli, Hosseini, and J{\'e}gou}]{douze2024faiss}
Matthijs Douze, Alexandr Guzhva, Chengqi Deng, Jeff Johnson, Gergely Szilvasy, Pierre-Emmanuel Mazar{\'e}, Maria Lomeli, Lucas Hosseini, and Herv{\'e} J{\'e}gou. 2024.
\newblock The faiss library.
\newblock \emph{arXiv preprint arXiv:2401.08281}.

\bibitem[{Enevoldsen et~al.(2025)Enevoldsen, Chung, Kerboua, Kardos, Mathur, Stap, Gala, Siblini, Krzemi{\'n}ski, Winata, Sturua, Utpala, Ciancone, Schaeffer, Misra, Dhakal, Rystr{\o}m, Solomatin, {\c{C}}a{\u{g}}atan, Kundu, Bernstorff, Xiao, Sukhlecha, Pahwa, Po{\'s}wiata, GV, Ashraf, Auras, Pl{\"u}ster, Harries, Magne, Mohr, Zhu, Gisserot-Boukhlef, Aarsen, Kostkan, Wojtasik, Lee, Suppa, Zhang, Rocca, Hamdy, Michail, Yang, Faysse, Vatolin, Thakur, Dey, Vasani, Chitale, Tedeschi, Tai, Snegirev, Hendriksen, G{\"u}nther, Xia, Shi, L{\`u}, Clive, K, Anna, Wehrli, Tikhonova, Panchal, Abramov, Ostendorff, Liu, Clematide, Miranda, Fenogenova, Song, Safi, Li, Borghini, Cassano, Hansen, Hooker, Xiao, Adlakha, Weller, Reddy, and Muennighoff}]{enevoldsen2025mmteb}
Kenneth Enevoldsen, Isaac Chung, Imene Kerboua, M{\'a}rton Kardos, Ashwin Mathur, David Stap, Jay Gala, Wissam Siblini, Dominik Krzemi{\'n}ski, Genta~Indra Winata, Saba Sturua, Saiteja Utpala, Mathieu Ciancone, Marion Schaeffer, Diganta Misra, Shreeya Dhakal, Jonathan Rystr{\o}m, Roman Solomatin, {\"O}mer~Veysel {\c{C}}a{\u{g}}atan, and 63 others. 2025.
\newblock \href {https://openreview.net/forum?id=zl3pfz4VCV} {{MMTEB}: Massive multilingual text embedding benchmark}.
\newblock In \emph{The Thirteenth International Conference on Learning Representations}.

\bibitem[{Grattafiori et~al.(2024)Grattafiori, Dubey, Jauhri, Pandey, Kadian, Al-Dahle, Letman, Mathur, Schelten, Vaughan et~al.}]{grattafiori2024llama}
Aaron Grattafiori, Abhimanyu Dubey, Abhinav Jauhri, Abhinav Pandey, Abhishek Kadian, Ahmad Al-Dahle, Aiesha Letman, Akhil Mathur, Alan Schelten, Alex Vaughan, and 1 others. 2024.
\newblock The llama 3 herd of models.
\newblock \emph{arXiv preprint arXiv:2407.21783}.

\bibitem[{Hu et~al.(2022)Hu, Shen, Wallis, Allen-Zhu, Li, Wang, Wang, Chen et~al.}]{hu2022lora}
Edward~J Hu, Yelong Shen, Phillip Wallis, Zeyuan Allen-Zhu, Yuanzhi Li, Shean Wang, Lu~Wang, Weizhu Chen, and 1 others. 2022.
\newblock Lora: Low-rank adaptation of large language models.
\newblock \emph{ICLR}, 1(2):3.

\bibitem[{Izacard et~al.(2021)Izacard, Caron, Hosseini, Riedel, Bojanowski, Joulin, and Grave}]{izacard2021unsupervised}
Gautier Izacard, Mathilde Caron, Lucas Hosseini, Sebastian Riedel, Piotr Bojanowski, Armand Joulin, and Edouard Grave. 2021.
\newblock Unsupervised dense information retrieval with contrastive learning.
\newblock \emph{arXiv preprint arXiv:2112.09118}.

\bibitem[{Johnson et~al.(2019)Johnson, Douze, and J{\'e}gou}]{johnson2019billion}
Jeff Johnson, Matthijs Douze, and Herv{\'e} J{\'e}gou. 2019.
\newblock Billion-scale similarity search with gpus.
\newblock \emph{IEEE Transactions on Big Data}, 7(3):535--547.

\bibitem[{Kwiatkowski et~al.(2019)Kwiatkowski, Palomaki, Redfield, Collins, Parikh, Alberti, Epstein, Polosukhin, Devlin, Lee et~al.}]{kwiatkowski2019natural}
Tom Kwiatkowski, Jennimaria Palomaki, Olivia Redfield, Michael Collins, Ankur Parikh, Chris Alberti, Danielle Epstein, Illia Polosukhin, Jacob Devlin, Kenton Lee, and 1 others. 2019.
\newblock Natural questions: a benchmark for question answering research.
\newblock \emph{Transactions of the Association for Computational Linguistics}, 7:453--466.

\bibitem[{Lan et~al.(2025)Lan, Chen, Liu, Li, Bao, and Lian}]{lan2025retro}
Junwei Lan, Jianlyu Chen, Zheng Liu, Chaofan Li, Siqi Bao, and Defu Lian. 2025.
\newblock Retro*: Optimizing llms for reasoning-intensive document retrieval.
\newblock \emph{arXiv preprint arXiv:2509.24869}.

\bibitem[{Lee et~al.(2024{\natexlab{a}})Lee, Roy, Xu, Raiman, Shoeybi, Catanzaro, and Ping}]{lee2024nvembed}
Chankyu Lee, Rajarshi Roy, Mengyao Xu, Jonathan Raiman, Mohammad Shoeybi, Bryan Catanzaro, and Wei Ping. 2024{\natexlab{a}}.
\newblock Nv-embed: Improved techniques for training llms as generalist embedding models.
\newblock \emph{arXiv preprint arXiv:2405.17428}.

\bibitem[{Lee et~al.(2025)Lee, Chen, Dua, Cer, Shanbhogue, Naim, {\'A}brego, Li, Chen, Vera et~al.}]{lee2025gemini}
Jinhyuk Lee, Feiyang Chen, Sahil Dua, Daniel Cer, Madhuri Shanbhogue, Iftekhar Naim, Gustavo~Hern{\'a}ndez {\'A}brego, Zhe Li, Kaifeng Chen, Henrique~Schechter Vera, and 1 others. 2025.
\newblock Gemini embedding: Generalizable embeddings from gemini.
\newblock \emph{arXiv preprint arXiv:2503.07891}.

\bibitem[{Lee et~al.(2024{\natexlab{b}})Lee, Dai, Ren, Chen, Cer, Cole, Hui, Boratko, Kapadia, Ding et~al.}]{lee2024gecko}
Jinhyuk Lee, Zhuyun Dai, Xiaoqi Ren, Blair Chen, Daniel Cer, Jeremy~R Cole, Kai Hui, Michael Boratko, Rajvi Kapadia, Wen Ding, and 1 others. 2024{\natexlab{b}}.
\newblock Gecko: Versatile text embeddings distilled from large language models.
\newblock \emph{arXiv preprint arXiv:2403.20327}.

\bibitem[{Li et~al.(2025{\natexlab{a}})Li, Chen, Shao, Lian, and Liu}]{li2025coder}
Chaofan Li, Jianlyu Chen, Yingxia Shao, Defu Lian, and Zheng Liu. 2025{\natexlab{a}}.
\newblock Towards a generalist code embedding model based on massive data synthesis.
\newblock \emph{arXiv preprint arXiv:2505.12697}.

\bibitem[{Li et~al.(2024)Li, Qin, Xiao, Chen, Luo, Shao, Lian, and Liu}]{li2024making}
Chaofan Li, MingHao Qin, Shitao Xiao, Jianlyu Chen, Kun Luo, Yingxia Shao, Defu Lian, and Zheng Liu. 2024.
\newblock Making text embedders few-shot learners.
\newblock \emph{arXiv preprint arXiv:2409.15700}.

\bibitem[{Li et~al.(2025{\natexlab{b}})Li, Zhou, and Liu}]{li2025r2med}
Lei Li, Xiao Zhou, and Zheng Liu. 2025{\natexlab{b}}.
\newblock R2med: A benchmark for reasoning-driven medical retrieval.
\newblock \emph{arXiv preprint arXiv:2505.14558}.

\bibitem[{Li et~al.(2023)Li, Zhang, Zhang, Long, Xie, and Zhang}]{li2023towards}
Zehan Li, Xin Zhang, Yanzhao Zhang, Dingkun Long, Pengjun Xie, and Meishan Zhang. 2023.
\newblock Towards general text embeddings with multi-stage contrastive learning.
\newblock \emph{arXiv preprint arXiv:2308.03281}.

\bibitem[{Liu et~al.(2025)Liu, Ma, Sun, Zhu, Li, Yin, and Dou}]{liu2025reasonrank}
Wenhan Liu, Xinyu Ma, Weiwei Sun, Yutao Zhu, Yuchen Li, Dawei Yin, and Zhicheng Dou. 2025.
\newblock Reasonrank: Empowering passage ranking with strong reasoning ability.
\newblock \emph{arXiv preprint arXiv:2508.07050}.

\bibitem[{Long et~al.(2025)Long, Sun, Yang, Wang, Shen, Wang, Wei, Gu, and Wang}]{long2025diver}
Meixiu Long, Duolin Sun, Dan Yang, Junjie Wang, Yue Shen, Jian Wang, Peng Wei, Jinjie Gu, and Jiahai Wang. 2025.
\newblock Diver: A multi-stage approach for reasoning-intensive information retrieval.
\newblock \emph{arXiv preprint arXiv:2508.07995}.

\bibitem[{Ma et~al.(2024)Ma, Wang, Yang, Wei, and Lin}]{ma2024repllama}
Xueguang Ma, Liang Wang, Nan Yang, Furu Wei, and Jimmy Lin. 2024.
\newblock Fine-tuning llama for multi-stage text retrieval.
\newblock In \emph{Proceedings of the 47th International ACM SIGIR Conference on Research and Development in Information Retrieval}, pages 2421--2425.

\bibitem[{Moreira et~al.(2024)Moreira, Osmulski, Xu, Ak, Schifferer, and Oldridge}]{moreira2024nv}
Gabriel de Souza~P Moreira, Radek Osmulski, Mengyao Xu, Ronay Ak, Benedikt Schifferer, and Even Oldridge. 2024.
\newblock Nv-retriever: Improving text embedding models with effective hard-negative mining.
\newblock \emph{arXiv preprint arXiv:2407.15831}.

\bibitem[{Muennighoff et~al.(2024)Muennighoff, Hongjin, Wang, Yang, Wei, Yu, Singh, and Kiela}]{muennighoff2024generative}
Niklas Muennighoff, SU~Hongjin, Liang Wang, Nan Yang, Furu Wei, Tao Yu, Amanpreet Singh, and Douwe Kiela. 2024.
\newblock Generative representational instruction tuning.
\newblock In \emph{The Thirteenth International Conference on Learning Representations}.

\bibitem[{Muennighoff et~al.(2022)Muennighoff, Tazi, Magne, and Reimers}]{muennighoff2022mteb}
Niklas Muennighoff, Nouamane Tazi, Lo{\"\i}c Magne, and Nils Reimers. 2022.
\newblock Mteb: Massive text embedding benchmark.
\newblock \emph{arXiv preprint arXiv:2210.07316}.

\bibitem[{Neelakantan et~al.(2022)Neelakantan, Xu, Puri, Radford, Han, Tworek, Yuan, Tezak, Kim, Hallacy et~al.}]{neelakantan2022text}
Arvind Neelakantan, Tao Xu, Raul Puri, Alec Radford, Jesse~Michael Han, Jerry Tworek, Qiming Yuan, Nikolas Tezak, Jong~Wook Kim, Chris Hallacy, and 1 others. 2022.
\newblock Text and code embeddings by contrastive pre-training.
\newblock \emph{arXiv preprint arXiv:2201.10005}.

\bibitem[{Niu et~al.(2024)Niu, Joty, Liu, Xiong, Zhou, and Yavuz}]{niu2024judgerank}
Tong Niu, Shafiq Joty, Ye~Liu, Caiming Xiong, Yingbo Zhou, and Semih Yavuz. 2024.
\newblock Judgerank: Leveraging large language models for reasoning-intensive reranking.
\newblock \emph{arXiv preprint arXiv:2411.00142}.

\bibitem[{Robertson et~al.(2009)Robertson, Zaragoza et~al.}]{robertson2009probabilistic}
Stephen Robertson, Hugo Zaragoza, and 1 others. 2009.
\newblock The probabilistic relevance framework: Bm25 and beyond.
\newblock \emph{Foundations and Trends{\textregistered} in Information Retrieval}, 3(4):333--389.

\bibitem[{Shao et~al.(2025)Shao, Qiao, Kishore, Muennighoff, Lin, Rus, Low, Min, Yih, Koh et~al.}]{shao2025reasonir}
Rulin Shao, Rui Qiao, Varsha Kishore, Niklas Muennighoff, Xi~Victoria Lin, Daniela Rus, Bryan Kian~Hsiang Low, Sewon Min, Wen-tau Yih, Pang~Wei Koh, and 1 others. 2025.
\newblock Reasonir: Training retrievers for reasoning tasks.
\newblock \emph{arXiv preprint arXiv:2504.20595}.

\bibitem[{Shinn et~al.(2023)Shinn, Cassano, Gopinath, Narasimhan, and Yao}]{shinn2023reflexion}
Noah Shinn, Federico Cassano, Ashwin Gopinath, Karthik Narasimhan, and Shunyu Yao. 2023.
\newblock Reflexion: Language agents with verbal reinforcement learning.
\newblock \emph{Advances in Neural Information Processing Systems}, 36:8634--8652.

\bibitem[{SU et~al.(2025)SU, Yen, Xia, Shi, Muennighoff, yu~Wang, Haisu, Shi, Siegel, Tang, Sun, Yoon, Arik, Chen, and Yu}]{su2025bright}
Hongjin SU, Howard Yen, Mengzhou Xia, Weijia Shi, Niklas Muennighoff, Han yu~Wang, Liu Haisu, Quan Shi, Zachary~S Siegel, Michael Tang, Ruoxi Sun, Jinsung Yoon, Sercan~O Arik, Danqi Chen, and Tao Yu. 2025.
\newblock \href {https://openreview.net/forum?id=ykuc5q381b} {{BRIGHT}: A realistic and challenging benchmark for reasoning-intensive retrieval}.
\newblock In \emph{The Thirteenth International Conference on Learning Representations}.

\bibitem[{Team(2024)}]{qwen2.5}
Qwen Team. 2024.
\newblock \href {https://qwenlm.github.io/blog/qwen2.5/} {Qwen2.5: A party of foundation models}.

\bibitem[{Vera et~al.(2025)Vera, Dua, Zhang, Salz, Mullins, Panyam, Smoot, Naim, Zou, Chen et~al.}]{vera2025embeddinggemma}
Henrique~Schechter Vera, Sahil Dua, Biao Zhang, Daniel Salz, Ryan Mullins, Sindhu~Raghuram Panyam, Sara Smoot, Iftekhar Naim, Joe Zou, Feiyang Chen, and 1 others. 2025.
\newblock Embeddinggemma: Powerful and lightweight text representations.
\newblock \emph{arXiv preprint arXiv:2509.20354}.

\bibitem[{Voorhees et~al.(2021)Voorhees, Alam, Bedrick, Demner-Fushman, Hersh, Lo, Roberts, Soboroff, and Wang}]{voorhees2021trec}
Ellen Voorhees, Tasmeer Alam, Steven Bedrick, Dina Demner-Fushman, William~R Hersh, Kyle Lo, Kirk Roberts, Ian Soboroff, and Lucy~Lu Wang. 2021.
\newblock Trec-covid: constructing a pandemic information retrieval test collection.
\newblock In \emph{ACM SIGIR Forum}, volume~54, pages 1--12. ACM New York, NY, USA.

\bibitem[{Wang et~al.(2024{\natexlab{a}})Wang, Ma, Feng, Zhang, Yang, Zhang, Chen, Tang, Chen, Lin et~al.}]{wang2024survey}
Lei Wang, Chen Ma, Xueyang Feng, Zeyu Zhang, Hao Yang, Jingsen Zhang, Zhiyuan Chen, Jiakai Tang, Xu~Chen, Yankai Lin, and 1 others. 2024{\natexlab{a}}.
\newblock A survey on large language model based autonomous agents.
\newblock \emph{Frontiers of Computer Science}, 18(6):186345.

\bibitem[{Wang et~al.(2022)Wang, Yang, Huang, Jiao, Yang, Jiang, Majumder, and Wei}]{wang2022text}
Liang Wang, Nan Yang, Xiaolong Huang, Binxing Jiao, Linjun Yang, Daxin Jiang, Rangan Majumder, and Furu Wei. 2022.
\newblock Text embeddings by weakly-supervised contrastive pre-training.
\newblock \emph{arXiv preprint arXiv:2212.03533}.

\bibitem[{Wang et~al.(2024{\natexlab{b}})Wang, Yang, Huang, Yang, Majumder, and Wei}]{wang2024e5mistral}
Liang Wang, Nan Yang, Xiaolong Huang, Linjun Yang, Rangan Majumder, and Furu Wei. 2024{\natexlab{b}}.
\newblock \href {https://doi.org/10.18653/v1/2024.acl-long.642} {Improving text embeddings with large language models}.
\newblock In \emph{Proceedings of the 62nd Annual Meeting of the Association for Computational Linguistics (Volume 1: Long Papers)}, pages 11897--11916, Bangkok, Thailand. Association for Computational Linguistics.

\bibitem[{Wang et~al.(2024{\natexlab{c}})Wang, Yang, Huang, Yang, Majumder, and Wei}]{wang2024multilingual}
Liang Wang, Nan Yang, Xiaolong Huang, Linjun Yang, Rangan Majumder, and Furu Wei. 2024{\natexlab{c}}.
\newblock Multilingual e5 text embeddings: A technical report.
\newblock \emph{arXiv preprint arXiv:2402.05672}.

\bibitem[{Weller et~al.(2025)Weller, Ricci, Yang, Yates, Lawrie, and Van~Durme}]{weller2025rank1}
Orion Weller, Kathryn Ricci, Eugene Yang, Andrew Yates, Dawn Lawrie, and Benjamin Van~Durme. 2025.
\newblock Rank1: Test-time compute for reranking in information retrieval.
\newblock \emph{arXiv preprint arXiv:2502.18418}.

\bibitem[{Xiao et~al.(2024)Xiao, Liu, Zhang, Muennighoff, Lian, and Nie}]{xiao2024c}
Shitao Xiao, Zheng Liu, Peitian Zhang, Niklas Muennighoff, Defu Lian, and Jian-Yun Nie. 2024.
\newblock C-pack: Packed resources for general chinese embeddings.
\newblock In \emph{Proceedings of the 47th international ACM SIGIR conference on research and development in information retrieval}, pages 641--649.

\bibitem[{Yang et~al.(2025{\natexlab{a}})Yang, Li, Yang, Zhang, Hui, Zheng, Yu, Gao, Huang, Lv et~al.}]{yang2025qwen3}
An~Yang, Anfeng Li, Baosong Yang, Beichen Zhang, Binyuan Hui, Bo~Zheng, Bowen Yu, Chang Gao, Chengen Huang, Chenxu Lv, and 1 others. 2025{\natexlab{a}}.
\newblock Qwen3 technical report.
\newblock \emph{arXiv preprint arXiv:2505.09388}.

\bibitem[{Yang et~al.(2025{\natexlab{b}})Yang, Yates, Ricci, Weller, Chari, Van~Durme, and Lawrie}]{yang2025rank}
Eugene Yang, Andrew Yates, Kathryn Ricci, Orion Weller, Vivek Chari, Benjamin Van~Durme, and Dawn Lawrie. 2025{\natexlab{b}}.
\newblock Rank-k: Test-time reasoning for listwise reranking.
\newblock \emph{arXiv preprint arXiv:2505.14432}.

\bibitem[{Yao et~al.(2023)Yao, Zhao, Yu, Du, Shafran, Narasimhan, and Cao}]{yao2023react}
Shunyu Yao, Jeffrey Zhao, Dian Yu, Nan Du, Izhak Shafran, Karthik Narasimhan, and Yuan Cao. 2023.
\newblock React: Synergizing reasoning and acting in language models.
\newblock In \emph{International Conference on Learning Representations (ICLR)}.

\bibitem[{Zhang et~al.(2024)Zhang, Li, Zeng, and Wang}]{zhang2024jasper}
Dun Zhang, Jiacheng Li, Ziyang Zeng, and Fulong Wang. 2024.
\newblock Jasper and stella: distillation of sota embedding models.
\newblock \emph{arXiv preprint arXiv:2412.19048}.

\bibitem[{Zhang et~al.(2025)Zhang, Li, Long, Zhang, Lin, Yang, Xie, Yang, Liu, Lin et~al.}]{zhang2025qwen3embed}
Yanzhao Zhang, Mingxin Li, Dingkun Long, Xin Zhang, Huan Lin, Baosong Yang, Pengjun Xie, An~Yang, Dayiheng Liu, Junyang Lin, and 1 others. 2025.
\newblock Qwen3 embedding: Advancing text embedding and reranking through foundation models.
\newblock \emph{arXiv preprint arXiv:2506.05176}.

\bibitem[{Zheng et~al.(2024{\natexlab{a}})Zheng, Yin, Xie, Sun, Huang, Yu, Cao, Kozyrakis, Stoica, Gonzalez et~al.}]{zheng2024sglang}
Lianmin Zheng, Liangsheng Yin, Zhiqiang Xie, Chuyue Sun, Jeff Huang, Cody~H Yu, Shiyi Cao, Christos Kozyrakis, Ion Stoica, Joseph~E Gonzalez, and 1 others. 2024{\natexlab{a}}.
\newblock Sglang: Efficient execution of structured language model programs.
\newblock \emph{Advances in neural information processing systems}, 37:62557--62583.

\bibitem[{Zheng et~al.(2024{\natexlab{b}})Zheng, Zhang, Zhang, Ye, Luo, Feng, and Ma}]{zheng2024llamafactory}
Yaowei Zheng, Richong Zhang, Junhao Zhang, Yanhan Ye, Zheyan Luo, Zhangchi Feng, and Yongqiang Ma. 2024{\natexlab{b}}.
\newblock Llamafactory: Unified efficient fine-tuning of 100+ language models.
\newblock \emph{arXiv preprint arXiv:2403.13372}.

\bibitem[{Zhuang et~al.(2025)Zhuang, Ma, Koopman, Lin, and Zuccon}]{zhuang2025rank}
Shengyao Zhuang, Xueguang Ma, Bevan Koopman, Jimmy Lin, and Guido Zuccon. 2025.
\newblock Rank-r1: Enhancing reasoning in llm-based document rerankers via reinforcement learning.
\newblock \emph{arXiv preprint arXiv:2503.06034}.

\end{thebibliography}

\newpage

\appendix

\section{Details on Data Synthesis Method}
\label{sec:appendix:data_synthesis}

In this section, we provide more details on our data synthesis method, \syn.

\subsection{Query Generation}

\textbf{Corpus Filtering Prompt}. Table~\ref{tab:corpus_filtering_prompt_tmpl} presents the prompt template used for filtering out domain-inconsistent documents from the entire corpus. We use the same LLM as used in query generation, Qwen2.5-72B-Instruct, to perform corpus filtering.

\textbf{Query Generation Prompt}. The designed prompt template for generating queries is presented in Table~\ref{tab:generation_prompt_tmpl}. The explicit generation instructions and output contents used in the query generation prompt are available in Table~\ref{tab:data_generation_instructions}. We refer to CodeR-Pile~\cite{li2025coder} and E5-Mistral~\cite{wang2024e5mistral} in the designing process of our prompt.

\textbf{Sampling Parameters}. For query generation, we set the temperature to 1.0. The inference is accelerated with SGLang~\cite{zheng2024sglang}.

\subsection{Candidate Mining}

\textbf{Retriever Implementation}. For a given task, we utilize gte-Qwen2-7B-instruct~\cite{li2023towards} to mine the top-100 most relevant documents from the corpus of this task as the candidate documents, where the Faiss~\cite{douze2024faiss} index is built and the ANN search method~\cite{johnson2019billion} is employed. The max query length and document length are both set to 8,192 tokens when encoding with gte-Qwen2-7B-instruct.

\textbf{Effectiveness Sensitivity to the Initial Retriever}. To explore the effectiveness sensitivity of our method to the initial retriever, we compare three different embedding models: 1) gte-Qwen2-7B-instruct (default setting), 2) Qwen3-Embedding-4B~\cite{zhang2025qwen3embed}, and 3) BGE-M3~\cite{chen2024m3}. For fair comparison, we maintain other settings the same as those used by default, including the relevance annotator, training settings, and evaluation settings. Table~\ref{tab:ablation:initial_retriever} shows their performance on BRIGHT to provide references for their retrieval capabilities in reasoning-intensive retrieval tasks. As observed, the initial retriever in our data synthesis workflow has a slight impact on the final performance of the fine-tuned embedding model. Notably, though BGE-M3 performs poorly on BRIGHT (nDCG@10 12.0), using it to mine candidate documents only leads to an nDCG@10 degradation of 2.2, which demonstrates the robustness of our data synthesis method.

\begin{table}[!t]
    \centering
    \small
    
    \setlength{\tabcolsep}{5pt}
        \begin{tabular}{l|c}
            \toprule
            \textbf{Model} & \textbf{Avg.} \\
            \midrule
            gte-Qwen2-7B-instruct               & 23.5 \\
            \quad Final \embed-Qwen3-8B    & 38.1 \\
            \midrule
            Qwen3-Embedding-4B                  & 21.8 \\
            \quad Final \embed-Qwen3-8B    & 37.2 \\
            \midrule
            BGE-M3                              & 12.0 \\
            \quad Final \embed-Qwen3-8B    & 35.9 \\
            \bottomrule
        \end{tabular}
    \caption{Impact of initial retriever (evaluated based on the \bright benchmark).}
    \label{tab:ablation:initial_retriever}
    \vspace{-5pt}
\end{table}

\subsection{Relevance Annotation}

\textbf{Relevance Annotation Prompt}. Table~\ref{tab:annotation_prompt_tmpl} presents the prompt template used for reasoning-enhanced relevance annotation. The crafted relevance definitions are available in Table~\ref{tab:relevance_annotation_instructions}.

\textbf{Distillation of Lightweight Annotator}. We first random sample 500 queries per task and sample 10 documents per query. We employ Qwen3-235B-A22B-Instruct-2507 to generate reasoning trajectories using the previously mentioned relevance annotation prompt for these sampled query-document pairs, leading to 60,000 training samples for distillation. We use the LLaMA-Factory framework~\cite{zheng2024llamafactory} to fine-tune Qwen3-8B on 8 NVIDIA H100 (80GB) GPUs in one epoch. The learning rate is set to 1e-5 and the warmup ratio is set to 0.1. The global training batch size is set to 8 and the gradient accumulation steps is set to 8.

\textbf{Sampling Parameters}. For relevance annotation, we set the temperature to 0.7. For Qwen3 series models, we disable thinking mode when using the tokenizer. The inference is accelerated with SGLang~\cite{zheng2024sglang}.

\subsection{Synthesization Result}

\textbf{Task Instructions}. The task instructions used in our synthetic dataset originate from \bright~\cite{su2025bright}, which are presented in Table~\ref{tab:bright_task_instructions}.

\textbf{Licenses}. The corpora in \bright used for data synthesis are licensed under CC BY-4.0\footnote{\url{https://choosealicense.com/licenses/cc-by-4.0/}}. our synthetic dataset is licensed under CC BY-NC-SA-4.0\footnote{\url{https://creativecommons.org/licenses/by-nc-sa/4.0}}. Therefore, our usage does not violate the intended use of \bright dataset.

\begin{figure}[!t]
  \centering
  \includegraphics[width=1.0\linewidth]{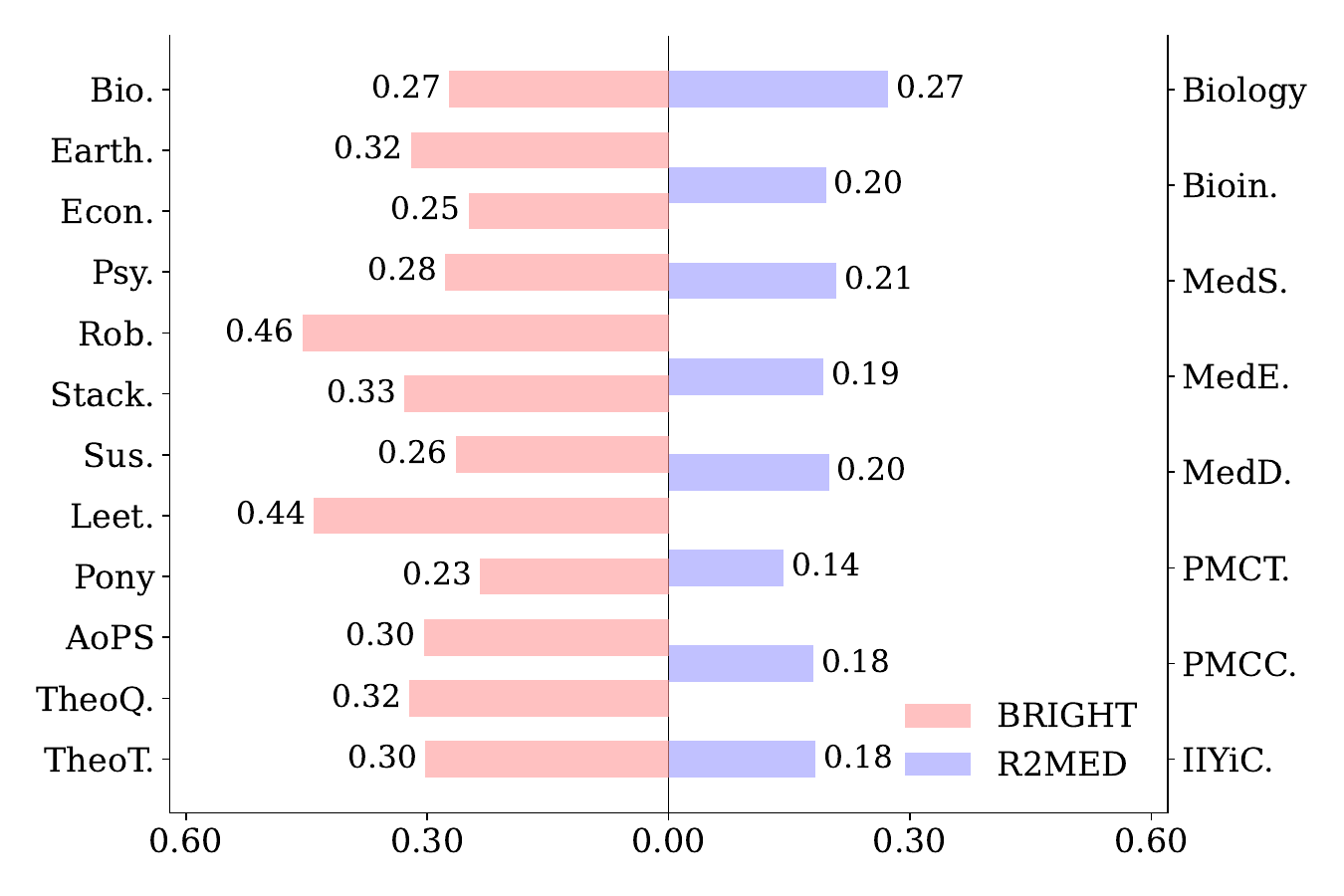}
  \caption{Data contamination analysis results (the computed max weighted Jaccard similarity) between our synthetic dataset and the testing data in \bright and \rrmed.}
  \label{fig:data_contamination_analysis}
  \vspace{-5pt}
\end{figure}

\textbf{Data Contamination Analysis}. To address the concerns of data leakage of the testing data in \bright and \rrmed, we perform a string-match-based query overlap analysis between the testing data and our synthetic dataset. Specifically, for each test query, we retrieve the top-20 most relevant training queries in our synthetic dataset\footnote{We only consider the corresponding training data in the same domain for \bright, and only consider the training data in the Biology domain for \rrmed.} using BM25~\cite{robertson2009probabilistic}, and then compute the weighted Jaccard similarity between the test query and each retrieved query. Figure~\ref{fig:data_contamination_analysis} presents the computed max similarity for each dataset, showcasing that there is no data leakage in our synthetic dataset.

\section{Details on Training Method}
\label{sec:appendix:training}

\subsection{Query Reasoning}

\textbf{Query Reasoning Prompt}. Table~\ref{tab:reasoning_query_prompt_tmpl} presents the prompt template used for generating reasoning queries.

\textbf{Sampling Parameters}. For generating reasoning-augmented queries, we set the temperature to 1.0. The max number of new tokens is set to 1024.

\subsection{Experimental Settings}
\label{sec:appendix:training_parameter}

\textbf{Hyperparameters}. The temperature used in Eq.~\ref{eq:info-nce} is set to 0.02. The truncating threshold used in Eq.~\ref{eq:reasoning_intensity} is set to 5.0. By default, we employ GPT-4.1-mini to generate the reasoning-augmented queries. 

\textbf{Query Instruction Template}. The query instruction template uses ``Instruct: \textit{\{task\_instruction\}}\textbackslash nQuery: \textit{\{query\}}''.

\textbf{Pooling Method}. In our experiments, the embeddings are obtained by taking the last layer [EOS] vector.

\textbf{Max Sequence Length}. The max sequence length of query and document during training are both set to 512.

\textbf{Negatives Strategy}. We use hard negatives, in-batch negatives and cross-device negatives during training, and the total number of negatives used for each query is 1023.

\textbf{Training}. For efficient fine-tuning, we employ Low-Rank Adaptation (LoRA)~\cite{hu2022lora} with LoRA rank set to 64 and LoRA alpha set to 32. The learning rate is set to 1e-4 and the warmup ratio is set to 0.1. The training is conducted on 8 NVIDIA H100 (80GB) GPUs with the FlagEmbedding\footnote{\url{https://github.com/FlagOpen/FlagEmbedding}} framework. The initialization processes on MSMARCO~\cite{bajaj2016ms} of Qwen3-4B-ms, Qwen3-8B-ms, Llama-3.1-8B-ms use the same training settings as listed above.

\subsection{Training Computation Cost Analysis}
\label{sec:appendix:training_cost_analysis}

Though our self-adaptive learning method \train involves computing reasoning intensity of samples during training, the training computation cost does not show significant change. As shown in Table~\ref{tab:ablation:training_cost_analysis}, when fine-tuning \embed-Qwen3-8B, the total train runtime using \train only increases 234 seconds compared to using the original contrastive learning method, demonstrating the efficiency of our self-adaptive learning method.

\begin{table}[!t]
    \centering
    \small
    
    \setlength{\tabcolsep}{5pt}
        \begin{tabular}{l|cc}
            \toprule
            \textbf{Method} & \textbf{\#Steps} & \textbf{Train Runtime / s} \\
            \midrule
            Contrastive Learning    & 1304 & 13144 \\
            \train (Ours)           & 1304 & 13378 \\
            \bottomrule
        \end{tabular}
    \caption{Training computation cost analysis of our self-adaptive learning method \train.}
    \label{tab:ablation:training_cost_analysis}
    \vspace{-5pt}
\end{table}

\section{Details on Evaluation}

\subsection{Baselines}
\label{appendix:experiment:baselines}

Table~\ref{tab:model_information} lists all of the baseline models shown in our paper.

\subsection{Evaluation Settings}

When performing evaluation on \bright, we use the task instructions presented in Table~\ref{tab:bright_task_instructions}, and set the max length of both query and document to 8,192 tokens. When performing evaluation on \rrmed, we use the task instructions presented in Table~\ref{tab:r2med_task_instructions}, and set the max length of both query and document to 512 tokens. We employ the evaluation framework from FlagEmbedding\footnote{\url{https://github.com/FlagOpen/FlagEmbedding}}.

\subsection{Detailed Evaluation Results of Ablation Study}
\label{appendix:experiment:detailed_ablation_results}

Table~\ref{tab:ablation:data_analysis_details} presents the detailed evaluation results of ablation study on candidate mining methods in Section~\ref{sec:ablation:data_synthesis}.
Table~\ref{tab:ablation:relevance_annotation_details} presents the detailed evaluation results of ablation study on relevance annotation methods in Section~\ref{sec:ablation:relevance_annotation}, where the annotation prompt template without reasoning used in \textsc{Non-Reason} is available in Table~\ref{tab:annotation_wo_reasoning_prompt_tmpl}.
Table~\ref{tab:ablation:training_details} presents the detailed evaluation results of ablation study on reasoning-intensity computation methods in Section~\ref{sec:ablation:training}.
Table~\ref{tab:ablation:training_data_size_details} presents the detailed evaluation results of ablation study on training data size scaling in Section~\ref{sec:ablation:data_size_scaling}.

\subsection{Evaluation on Non-Reasoning-intensive Retrieval Tasks}
\label{appendix:non_reasoning_intensive_performance}

To demonstrate the effectiveness of our synthetic data on non-reasoning-intensive retrieval tasks, we fine-tune Qwen3-8B using our synthetic data and NQ~\cite{kwiatkowski2019natural}, respectively, and then perform evaluation on two non-reasoning-intensive tasks, including TREC-COVID~\cite{voorhees2021trec} and KAPR~\cite{capari2024knowledge}. Table~\ref{tab:non_reasoning_intensive_results} presents the performances on TREC-COVID and KAPR using different training data to fine-tune the Qwen3-8B base model. We can make the following observations: 1) First, though TREC-COVID and KAPR are not reasoning-intensive retrieval tasks, our synthetic data still helps to improve the model's retrieval performance on them. Specifically, performances on TREC-COVID and KAPR consistently improve as we scale the size of our synthetic dataset, demonstrating the generalizability of its effectiveness. 2) Second, our data achieves an even better impact on traditional tasks than NQ, a popular dataset for traditional question-answering retrieval, despite the fact that it focuses on reasoning-intensive scenarios. We believe the results here provide strong evidence about the effectiveness of our data synthesis method.

\begin{table}[!t]
    \centering
    \small
    
    \setlength{\tabcolsep}{5pt}
    \setlength{\extrarowheight}{2pt}
    \resizebox{0.48\textwidth}{!}{
        \begin{tabular}{lc|cc}
            \toprule
            \textbf{Training Data}  & \#Samples & TREC-COVID & KAPR \\
            \midrule
            NQ              & 58,568 & 64.40 & 75.26 \\
            Ours (10K)      & 10,202 & 57.68 & 54.58 \\
            Ours (20K)      & 20,411 & 75.20 & 63.93 \\
            Ours (full 80K) & 81,659 & 79.86 & 76.99 \\
            \bottomrule
        \end{tabular}
    }
    \caption{Performances on TREC-COVID and KAPR using different training data to fine-tune Qwen3-8B base model.}
    \label{tab:non_reasoning_intensive_results}
    \vspace{-5pt}
\end{table}

\begin{table}[!ht]
    \centering
    
    \small{\begin{tabular}{l}
    \toprule
    \begin{tabular}[c]{@{}p{0.95\linewidth}@{}}
    Given a passage, determine whether it belongs to the domain: \textit{\{Domain\}}\\ \\
    
    The given passage: \\
    {[}Begin of Passage{]} \\
    \textit{\{Doc\}} \\
    {[}End of Passage{]} \\ \\
    
    Note: \\
    - Your output must always be ``Yes'' or ``No''. \\ \\
    
    Remember do not explain your output or output anything else. Your output:\end{tabular} \\
    \bottomrule
    \end{tabular}}
    \caption{Prompt template for filtering our domain-inconsistent documents from the raw corpus. The placeholder \textit{\{Domain\}} is set by the following mapping function: \textit{\{Biology, Earth Science, Economics, Psychology, Robotics, Sustainable Living\}} $\rightarrow$ \textit{the original dataset name}. \textit{\{Stack Overflow, LeetCode, Pony\}} $\rightarrow$ \textit{Coding}. \textit{\{AoPS, TheoremQA Questions, TheoremQA Theorems\}} $\rightarrow$ \textit{Math}.}
    \label{tab:corpus_filtering_prompt_tmpl}
    \vspace{-5pt}
\end{table}

\newpage

\begin{table}[!ht]
    \centering
    
    \small{\begin{tabular}{l}
    \toprule
    \begin{tabular}[c]{@{}p{0.97\linewidth}@{}}
    \textit{\{Generation Instruction\}}\\ \\
    
    The given content: \\
    {[}Begin of Content{]} \\
    \textit{\{Input Content\}} \\
    {[}End of Content{]} \\ \\
    
    Note: \\
    - Your output must always be a string, only containing \textit{\{Output Content\}}. \\
    - Your output should be independent of the given content, which means that it should not containing the pronouns such as "it", "this", "that", "the given", "the provided", etc. \\
    - Your output (\textit{\{Output Content\}}) should be about \textit{\{Length\}}. \\
    - Your output (\textit{\{Output Content\}}) should require \textit{\{Difficulty\}} level education to understand. \\ \\
    
    Remember do not explain your output or output anything else. Your output:\end{tabular} \\
    \bottomrule
    \end{tabular}}
    \caption{Prompt template for generating task-consistent queries. For placeholders, ``\textit{\{Generation Instruction\}}'' $\in$ Table~\ref{tab:data_generation_instructions}, ``\textit{\{Output Content\}}'' $\in$ Table~\ref{tab:data_generation_instructions}, ``\textit{\{Length\}}'' $\in$ \{less than 100 words, 100 to 200 words, 200 to 300 words, 300 to 400 words, 400 to 500 words, at least 500 words\}, and ``\textit{\{Difficulty\}}'' $\in$ \{high school, college, phd\}.}
    \label{tab:generation_prompt_tmpl}
    \vspace{-5pt}
\end{table}

\begin{table}[!ht]
    \centering
    
    \small{\begin{tabular}{l}
    \toprule
    \begin{tabular}[c]{@{}p{0.97\linewidth}@{}}
    Given a question, your mission is to follow the instructions below: \\
    1. Identify the essential problem. \\
    2. Think step by step to reason and describe what information could be relevant and helpful to address the questions in detail. \\
    3. Draft an answer with as many thoughts as you have. \\ \\
    
    The given question: \\
    {[}Begin of Question{]} \\
    \textit{\{Original Query\}} \\
    {[}End of Question{]} \end{tabular} \\
    \bottomrule
    \end{tabular}}
    \caption{Prompt template for generating reasoning queries for original queries.}
    \label{tab:reasoning_query_prompt_tmpl}
    \vspace{-5pt}
\end{table}

\newpage

\begin{table}[!ht]
    \centering
    
    \small{\begin{tabular}{l}
    \toprule
    \begin{tabular}[c]{@{}p{0.97\linewidth}@{}}
    Here is the \textbf{relevance definition} in a retrieval task: \textit{\{Relevance Definition\}}\\ \\
    
    Now given a \textbf{query} (\textit{\{Query Type\}}) and a \textbf{document} (\textit{\{Doc Type\}}) in this retrieval task, your mission is to perform the following steps to determine the relevance between the query and the document. \\ \\

    1. Query Analysis: Think to reason and describe what information would be most helpful in answering the query. \\
    2. Document Analysis: Discuss how the information provided by the document fulfills or fails to fulfill the requirements implied by the query. \\
    3. Relevance Annotation: Based on the relevance definition and the insights from the previous two steps, clearly justify your final relevance annotation result and annotate an integer score from a scale of 1 to 5. Please use the following guide: \\
    \quad - \textbf{5 (Highly Relevant):} The document is directly and fully responsive to the query, providing comprehensive, accurate, and specific information that completely addresses all aspects of the query.\\
    \quad - \textbf{4 (Relevant):} The document is largely relevant and provides most of the information needed, but may have minor omissions, slight inaccuracies, or not be perfectly aligned with the query's intent.\\
    \quad - \textbf{3 (Moderately Relevant):} The document has some relevance and offers partial information, but it may be incomplete, vague, or include some irrelevant content. It provides a basic connection but lacks depth or precision.\\
    \quad - \textbf{2 (Slightly Relevant):} The document has minimal relevance, with only a small portion of content tangentially related to the query. The majority of the document is off-topic or provides little value.\\
    \quad - \textbf{1 (Irrelevant):} The document is completely unrelated to the query and provides no useful information. There is no discernible connection or value for answering the query.\\ \\
    
    After providing your detailed analysis and justification for all the steps above, conclude your entire response with the final relevance score. The score must be placed strictly between the <score> tags. There should be no other text or explanation inside the tags:\\
    <score>\\
    {[}From a scale of 1 to 5, annotate the degree of relevance between the query and the document.{]}\\
    </score>\\ \\
    
    Note: The whole response should be as concise as possible while covering all the necessary details, and not exceeding 512 words in total.\\ \\

    Query (\textit{\{Query Type\}}): \\
    {[}Begin of Query{]} \\
    \textit{\{Query\}} \\
    {[}End of Query{]} \\ \\
    
    Document (\textit{\{Doc Type\}}): \\
    {[}Begin of Document{]} \\
    \textit{\{Doc\}} \\
    {[}End of Document{]} \\
    \end{tabular} \\
    \bottomrule
    \end{tabular}}
    \caption{Prompt template for annotating the relevance of query-document pair. For placeholders, ``\textit{\{Query Type\}}'', ``\textit{\{Doc Type\}}'', ``\textit{\{Relevance Definition\}}'' $\in$ Table~\ref{tab:relevance_annotation_instructions}.}
    \label{tab:annotation_prompt_tmpl}
    \vspace{-5pt}
\end{table}

\begin{table}[!ht]
    \centering

    \resizebox{0.48\textwidth}{!}{
        \begin{tabular}{ll}
        \toprule
        \textbf{Task Name} & \textbf{Task Instruction} \\
        \midrule
        \multicolumn{2}{l}{\textit{\textbf{StackExchange (7)}}} \\
        \midrule
        Bio.             & \begin{tabular}[c]{@{}l@{}} Given a Biology post, retrieve relevant \\ passages that help answer the post. \end{tabular} \\
        Earth.       & \begin{tabular}[c]{@{}l@{}} Given an Earth Science post, retrieve relevant \\ passages that help answer the post. \end{tabular} \\
        Econ.           & \begin{tabular}[c]{@{}l@{}} Given an Economics post, retrieve relevant \\ passages that help answer the post. \end{tabular} \\
        Psy.          & \begin{tabular}[c]{@{}l@{}} Given a Psychology post, retrieve relevant \\ passages that help answer the post. \end{tabular} \\
        Rob.            & \begin{tabular}[c]{@{}l@{}} Given a Robotics post, retrieve relevant \\ passages that help answer the post. \end{tabular} \\
        Stack.      & \begin{tabular}[c]{@{}l@{}} Given a Stack Overflow post, retrieve relevant \\ passages that help answer the post. \end{tabular} \\
        Sus.  & \begin{tabular}[c]{@{}l@{}} Given a Sustainable Living post, retrieve \\ relevant passages that help answer the post. \end{tabular} \\
        \midrule
        \multicolumn{2}{l}{\textit{\textbf{Coding (2)}}} \\
        \midrule
        Leet.            & \begin{tabular}[c]{@{}l@{}} Given a Coding problem, retrieve relevant \\ examples that help answer the problem. \end{tabular} \\
        Pony                & \begin{tabular}[c]{@{}l@{}} Given a Pony question, retrieve relevant \\ passages that help answer the question. \end{tabular} \\
        \midrule
        \multicolumn{2}{l}{\textit{\textbf{Math (3)}}} \\
        \midrule
        AoPS                & \begin{tabular}[c]{@{}l@{}} Given a Math problem, retrieve relevant \\ examples that help answer the problem. \end{tabular} \\
        TheoQ. & \begin{tabular}[c]{@{}l@{}} Given a Math problem, retrieve relevant \\ examples that help answer the problem. \end{tabular} \\
        TheoT.  & \begin{tabular}[c]{@{}l@{}} Given a Math problem, retrieve relevant \\ theorems that help answer the problem. \end{tabular} \\
        \bottomrule
        \end{tabular}
    }
    \caption{Task names and task instructions for all 12 retrieval tasks in our synthetic dataset.}
    \label{tab:bright_task_instructions}
    \vspace{-5pt}
\end{table}

\begin{table}[!ht]
    \centering

    \resizebox{0.48\textwidth}{!}{
        \begin{tabular}{ll}
        \toprule
        \textbf{Task Name} & \textbf{Task Instruction} \\
        \midrule
        \multicolumn{2}{l}{\textit{\textbf{Q\&A Reference (3)}}} \\
        \midrule
        Biology             & \begin{tabular}[c]{@{}l@{}} Given a Biology post, retrieve relevant \\ passages that help answer the post. \end{tabular} \\
        Bioin.       & \begin{tabular}[c]{@{}l@{}} Given a Bioinformatics post, retrieve relevant \\ passages that help answer the post. \end{tabular} \\
        MedS.           & \begin{tabular}[c]{@{}l@{}} Given a Medical Science post, retrieve relevant \\ passages that help answer the post. \end{tabular} \\
        \midrule
        \multicolumn{2}{l}{\textit{\textbf{Clinical Evidence (3)}}} \\
        \midrule
        MedE.            & \begin{tabular}[c]{@{}l@{}} Given a Medical Exam, retrieve relevant \\ passages that help answer the exam. \end{tabular} \\
        MedD.                & \begin{tabular}[c]{@{}l@{}} Given a Medical Exam, retrieve relevant \\ passages that help answer the exam. \end{tabular} \\
        PMCT.                & \begin{tabular}[c]{@{}l@{}} Given a Clinical Case, retrieve relevant \\ passages that help answer the case. \end{tabular} \\
        \midrule
        \multicolumn{2}{l}{\textit{\textbf{Clinical Case (2)}}} \\
        \midrule
        PMCC. & \begin{tabular}[c]{@{}l@{}} Given a Clinical Case, retrieve similar cases \\ that help diagnose the case. \end{tabular} \\
        IIYiC.  & \begin{tabular}[c]{@{}l@{}} Given a Clinical Case, retrieve similar cases \\ that help diagnose the case. \end{tabular} \\
        \bottomrule
        \end{tabular}
    }
    \caption{Task names and task instructions for all 8 retrieval tasks in \rrmed.}
    \label{tab:r2med_task_instructions}
    \vspace{-5pt}
\end{table}

\begin{table}[!ht]
    \centering
    
    \small{\begin{tabular}{l}
    \toprule
    \begin{tabular}[c]{@{}p{0.97\linewidth}@{}}
    Here is the \textbf{relevance definition} in a retrieval task: \textit{\{Relevance Definition\}}\\ \\
    
    Now given a \textbf{query} (\textit{\{Query Type\}}) and a \textbf{document} (\textit{\{Doc Type\}}) in this retrieval task, your mission is to perform the following steps to determine the relevance between the query and the document. \\ \\

    1. Query Analysis: Think to reason and describe what information would be most helpful in answering the query. \\
    2. Document Analysis: Discuss how the information provided by the document fulfills or fails to fulfill the requirements implied by the query. \\
    3. Relevance Annotation: Based on the relevance definition and the insights from the previous two steps, clearly justify your final relevance annotation result and annotate an integer score from a scale of 1 to 5. Please use the following guide: \\
    \quad - \textbf{5 (Highly Relevant):} The document is directly and fully responsive to the query, providing comprehensive, accurate, and specific information that completely addresses all aspects of the query.\\
    \quad - \textbf{4 (Relevant):} The document is largely relevant and provides most of the information needed, but may have minor omissions, slight inaccuracies, or not be perfectly aligned with the query's intent.\\
    \quad - \textbf{3 (Moderately Relevant):} The document has some relevance and offers partial information, but it may be incomplete, vague, or include some irrelevant content. It provides a basic connection but lacks depth or precision.\\
    \quad - \textbf{2 (Slightly Relevant):} The document has minimal relevance, with only a small portion of content tangentially related to the query. The majority of the document is off-topic or provides little value.\\
    \quad - \textbf{1 (Irrelevant):} The document is completely unrelated to the query and provides no useful information. There is no discernible connection or value for answering the query.\\ \\
    
    Directly output the final relevance score without any explanation or reasoning steps. The score must be placed strictly between the <score> tags. There should be no other text or explanation inside the tags:\\
    <score>\\
    {[}From a scale of 1 to 5, annotate the degree of relevance between the query and the document.{]}\\
    </score>\\ \\
    
    Note: The response should ONLY contain the score enclosed within the <score> tags, with no additional text or commentary. Example of correct format: <score>4</score>. \\ \\

    Query (\textit{\{Query Type\}}): \\
    {[}Begin of Query{]} \\
    \textit{\{Query\}} \\
    {[}End of Query{]} \\ \\
    
    Document (\textit{\{Doc Type\}}): \\
    {[}Begin of Document{]} \\
    \textit{\{Doc\}} \\
    {[}End of Document{]} \\
    \end{tabular} \\
    \bottomrule
    \end{tabular}}
    \caption{Prompt template for annotating the relevance of query-document pair \textbf{without reasoning process} (used for ablation study in Section~\ref{sec:ablation:relevance_annotation}). For placeholders, ``\textit{\{Query Type\}}'', ``\textit{\{Doc Type\}}'', ``\textit{\{Relevance Definition\}}'' $\in$ Table~\ref{tab:relevance_annotation_instructions}.}
    \label{tab:annotation_wo_reasoning_prompt_tmpl}
    \vspace{-5pt}
\end{table}

\newpage

\begin{table*}[!ht]
    \centering

    \resizebox{\textwidth}{!}{
        \begin{tabular}{lll}
        \toprule
        \textbf{Task Name} & \textbf{Generation Instruction} & \textbf{Output Content} \\ 
        \midrule
        \multicolumn{3}{l}{\textit{\textbf{StackExchange (7)}}} \\
        \midrule
        Bio.             & \begin{tabular}[c]{@{}l@{}} Given a Biology-related passage in \textit{\{language\}}, generate a StackExchange post in \textit{\{language\}} for which the critical \\ concepts or theories discussed in the passage can serve as references for domain experts to draft an answer. \end{tabular} & the generated StackExchange post in \textit{\{language\}} \\
        Earth.       & \begin{tabular}[c]{@{}l@{}} Given a Biology-related passage in \textit{\{language\}}, generate a StackExchange post in \textit{\{language\}} for which the critical \\ concepts or theories discussed in the passage can serve as references for domain experts to draft an answer. \end{tabular} & the generated StackExchange post in \textit{\{language\}} \\
        Econ.           & \begin{tabular}[c]{@{}l@{}} Given an Economics-related passage in \textit{\{language\}}, generate a StackExchange post in \textit{\{language\}} for which the critical \\ concepts or theories discussed in the passage can serve as references for domain experts to draft an answer. \end{tabular} & the generated StackExchange post in \textit{\{language\}} \\
        Psy.          & \begin{tabular}[c]{@{}l@{}} Given a Psychology-related passage in \textit{\{language\}}, generate a StackExchange post in \textit{\{language\}} for which the critical \\ concepts or theories discussed in the passage can serve as references for domain experts to draft an answer. \end{tabular} & the generated StackExchange post in \textit{\{language\}} \\
        Rob.            & \begin{tabular}[c]{@{}l@{}} Given a Robotics-related passage in \textit{\{language\}}, generate a StackExchange post in \textit{\{language\}} for which the critical \\ concepts or theories discussed in the passage can serve as references for domain experts to draft an answer. \end{tabular} & the generated StackExchange post in \textit{\{language\}} \\
        Stack.      & \begin{tabular}[c]{@{}l@{}} Given a Coding-related passage in \textit{\{language\}}, generate a StackExchange post in \textit{\{language\}} for which the critical \\ concepts or theories discussed in the passage can serve as references for domain experts to draft an answer. \end{tabular} & the generated StackExchange post in \textit{\{language\}} \\
        Sus.  & \begin{tabular}[c]{@{}l@{}} Given a Sustainable Living-related passage in \textit{\{language\}}, generate a StackExchange post in \textit{\{language\}} for which the \\ critical concepts or theories discussed in the passage can serve as references for domain experts to draft an answer. \end{tabular} & the generated StackExchange post in \textit{\{language\}} \\
        \midrule
        \multicolumn{3}{l}{\textit{\textbf{Coding (2)}}} \\
        \midrule
        Leet.            & \begin{tabular}[c]{@{}l@{}} Given a solved LeetCode problem (with solutions) in \textit{\{language\}}, generate a new LeetCode problem in \textit{\{language\}} \\ that the underlying algorithms or data structures from the original problem can help solve. \end{tabular} & the generated LeetCode problem in \textit{\{language\}} \\
        Pony                & \begin{tabular}[c]{@{}l@{}} Given a Pony documentation passage in \textit{\{language\}}, generate a Pony coding instruction in \textit{\{language\}} that the Pony \\ syntax described in the passage can help implement. \end{tabular} & the generated Pony coding instruction in \textit{\{language\}} \\
        \midrule
        \multicolumn{3}{l}{\textit{\textbf{Math (3)}}} \\
        \midrule
        AoPS                & \begin{tabular}[c]{@{}l@{}} Given a Math problem solution in \textit{\{language\}}, generate a new Math problem in \textit{\{language\}} that the problem-solving \\ skills used in the original problem can help solve. \end{tabular} & the generated Math problem in \textit{\{language\}} \\
        TheoQ. & \begin{tabular}[c]{@{}l@{}} Given a Math problem solution in \textit{\{language\}}, generate a new Math problem in \textit{\{language\}} that the theorems used in \\ the original problem can help solve. \end{tabular} & the generated Math problem in \textit{\{language\}} \\
        TheoT.  & \begin{tabular}[c]{@{}l@{}} Given a Math theorem in \textit{\{language\}}, generate a Math problem in \textit{\{language\}} that the theorem can help solve. \end{tabular} & the generated Math problem in \textit{\{language\}} \\
        \bottomrule
        \end{tabular}
    }
    \caption{Generation instructions and output contents for all 12 retrieval tasks in our synthetic dataset. The placeholder \textit{\{language\}} is set to \textit{English}.}
    \label{tab:data_generation_instructions}
    \vspace{-5pt}
\end{table*}

\begin{table*}[!ht]
    \centering

    \resizebox{\textwidth}{!}{
        \begin{tabular}{ll}
        \toprule
        \textbf{Task Name} & \textbf{Generation Instruction} \\ 
        \midrule
        \multicolumn{2}{l}{\textit{\textbf{StackExchange (7)}}} \\
        \midrule
        Bio.             & \begin{tabular}[c]{@{}l@{}} Given a query (\textit{biology post}) and a document (\underline{passage}), the document is relevant to the query if the critical concepts or theories discussed in the \\ document can provide references for domain experts to draft an answer to the query. \end{tabular} \\
        Earth.       & \begin{tabular}[c]{@{}l@{}} Given a query (\textit{earth science post}) and a document (\underline{passage}), the document is relevant to the query if the critical concepts or theories discussed in \\ the document can provide references for domain experts to draft an answer to the query. \end{tabular} \\
        Econ.           & \begin{tabular}[c]{@{}l@{}} Given a query (\textit{economics post}) and a document (\underline{passage}), the document is relevant to the query if the critical concepts or theories discussed in \\ the document can provide references for domain experts to draft an answer to the query. \end{tabular} \\
        Psy.          & \begin{tabular}[c]{@{}l@{}} Given a query (\textit{psychology post}) and a document (\underline{passage}), the document is relevant to the query if the critical concepts or theories discussed in \\ the document can provide references for domain experts to draft an answer to the query. \end{tabular} \\
        Rob.            & \begin{tabular}[c]{@{}l@{}} Given a query (\textit{robotics post}) and a document (\underline{passage}), the document is relevant to the query if the critical concepts or theories discussed in the \\ document can provide references for domain experts to draft an answer to the query. \end{tabular} \\
        Stack.      & \begin{tabular}[c]{@{}l@{}} Given a query (\textit{Stack Overflow post}) and a document (\underline{passage}), the document is relevant to the query if the critical concepts or theories discussed \\ in the document can provide references for domain experts to draft an answer to the query. \end{tabular} \\
        Sus.  & \begin{tabular}[c]{@{}l@{}} Given a query (\textit{sustainable living post}) and a document (\underline{passage}), the document is relevant to the query if the critical concepts or theories discussed \\ in the document can provide references for domain experts to draft an answer to the query. \end{tabular} \\
        \midrule
        \multicolumn{2}{l}{\textit{\textbf{Coding (2)}}} \\
        \midrule
        Leet.            & \begin{tabular}[c]{@{}l@{}} Given a query (\textit{LeetCode problem}) and a document (\underline{coding problem solution}), the document is relevant to the query if the underlying algorithms \\ or data structures used in the document can provide helpful insights for solving the problem in the query. \end{tabular} \\
        Pony                & \begin{tabular}[c]{@{}l@{}} Given a query (\textit{Pony coding instruction}) and a document (\underline{Pony documentation passage}), the document is relevant to the query if the Pony syntax \\ described in the document is necessary for beginners with no prior knowledge of Pony to complete the coding instruction in the query. \end{tabular} \\
        \midrule
        \multicolumn{2}{l}{\textit{\textbf{Math (3)}}} \\
        \midrule
        AoPS                & \begin{tabular}[c]{@{}l@{}} Given a query (\textit{math problem}) and a document (\underline{math problem solution}), the document is relevant to the query if the theorems used in the document \\ can provide helpful insights for solving the problem in the query. \end{tabular} \\
        TheoQ. & \begin{tabular}[c]{@{}l@{}} Given a query (\textit{math problem}) and a document (\underline{math problem solution}), the document is relevant to the query if the theorems used in the document \\ can provide helpful insights for solving the problem in the query. \end{tabular} \\
        TheoT.  & \begin{tabular}[c]{@{}l@{}} Given a query (\textit{math problem}) and a document (\underline{math-related passage}), the document is relevant to the query if the theorem described in the \\ document can help solve the problem in the query. \end{tabular} \\
        \bottomrule
        \end{tabular}
    }
    \caption{Relevance definitions used for annotation for all 12 retrieval tasks in our synthetic dataset. The \textit{query type} is italic, and the \underline{document type} is underlined.}
    \label{tab:relevance_annotation_instructions}
    \vspace{-5pt}
\end{table*}

\begin{table*}[!ht]
    \centering
    \small
    \resizebox{\textwidth}{!}{
        \begin{tabular}{l|c|c}
        \toprule
        \textbf{Model} & \textbf{Size} & \textbf{Model Link} \\
        \midrule
        \multicolumn{3}{l}{\textit{Lexical method}} \\
        \midrule
        BM25~\cite{robertson2009probabilistic} & - & \url{https://github.com/xlang-ai/BRIGHT} \\
        \midrule
        \multicolumn{3}{l}{\textit{General-purpose embedding models}} \\
        \midrule
        OpenAI-3-Large & - & \url{https://openai.com/index/new-embedding-models-and-api-updates} \\
        Google-Gecko-1B-768~\cite{lee2024gecko} & 1B & \url{https://cloud.google.com/vertex-ai/generative-ai/docs/model-reference/text-embeddings-api} \\
        GritLM-7B~\cite{muennighoff2024generative} & 7B & \url{https://huggingface.co/GritLM/GritLM-7B} \\
        NV-Embed-v2~\cite{lee2024nvembed} & 7B & \url{https://huggingface.co/nvidia/NV-Embed-v2} \\
        gte-Qwen2-7B-instruct~\cite{li2023towards} & 7B & \url{https://huggingface.co/Alibaba-NLP/gte-Qwen2-7B-instruct} \\
        Qwen3-Embedding-4B~\cite{zhang2025qwen3embed} & 4B & \url{https://huggingface.co/Qwen/Qwen3-Embedding-4B} \\
        Qwen3-Embedding-8B~\cite{zhang2025qwen3embed} & 8B & \url{https://huggingface.co/Qwen/Qwen3-Embedding-8B} \\
        \midrule
        \multicolumn{3}{l}{\textit{Reasoning-optimized embedding models}} \\
        \midrule
        ReasonIR-8B~\cite{shao2025reasonir} & 8B & \url{https://huggingface.co/reasonir/ReasonIR-8B} \\
        RaDeR-gte-Qwen2-7B~\cite{das2025rader} & 7B & \url{https://huggingface.co/Raderspace/RaDeR_gte_Qwen2-7B_MATH_LLMq_CoT_lexical} \\
        Seed-1.5-Embedding & - & \url{https://seed1-5-embedding.github.io/} \\
        DIVER-Retriever~\cite{long2025diver} & 4B & \url{https://huggingface.co/AQ-MedAI/Diver-Retriever-4B} \\
        \bottomrule
    \end{tabular}
    }
    \caption{Detailed information on all of the baseline models shown in our paper.}
    \label{tab:model_information}
    \vspace{-5pt}
\end{table*}

\begin{table*}[!ht]
    \centering
    
    \setlength{\tabcolsep}{3pt}
    \resizebox{1.0\textwidth}{!}{
        \begin{tabular}{l|c|c|c|ccccccc|cc|ccc}
            \toprule
            \multirow{2}{*}{\textbf{Method}} & \multirow{2}{*}{\textbf{Positive}} & \multirow{2}{*}{\textbf{Negative}} & \multirow{2}{*}{\textbf{Avg.}} & \multicolumn{7}{c|}{\textbf{StackExchange}} & \multicolumn{2}{c|}{\textbf{Coding}} & \multicolumn{3}{c}{\textbf{Theorem-based}} \\
            \cmidrule{5-16}
            & & & & \textbf{Bio.} & \textbf{Earth.} & \textbf{Econ.} & \textbf{Psy.} & \textbf{Rob.} & \textbf{Stack.} & \textbf{Sus.} & \textbf{Leet.} & \textbf{Pony} & \textbf{AoPS} & \textbf{TheoQ.} & \textbf{TheoT.} \\
            \midrule
            \multicolumn{16}{l}{\textbf{{\embed-Qwen3-8B from basic contrastive learning (using InfoNCE loss)}}} \\
            \midrule
            \underline{\textsc{Default}}    & mined \& annotat. & mined \& annotat.     & 37.1 & 54.4 & 55.4 & 33.8 & 45.2 & 32.0 & 34.3 & 37.3 & 32.3 & 18.7 & 13.3 & 41.2 & 47.6 \\
            \underline{\textsc{Non-Anno}}   & source            & mined                 & 16.1 & 13.1 & 26.2 & 11.1 & 14.6 & 15.1 & 16.3 & 14.5 & 28.8 & 5.0  & 6.2  & 22.8 & 19.0 \\
            \underline{\textsc{Source}}     & source            & mined \& annotat.     & 14.5 & 11.4 & 22.0 & 12.9 & 12.6 & 15.6 & 16.3 & 15.5 & 23.5 & 5.0  & 6.4  & 20.4 & 12.4 \\
            \underline{\textsc{Source-Pro}} & source \& annotat.& mined \& annotat.     & 22.1 & 28.0 & 38.0 & 19.3 & 20.8 & 18.4 & 23.0 & 19.3 & 29.6 & 4.3  & 6.5  & 27.1 & 30.4 \\
            \bottomrule
        \end{tabular}
    }
    \caption{Detailed evaluation results (nDCG@10) on \bright benchmark (using original queries) for ablation study of \textbf{candidate mining} methods.}
    \label{tab:ablation:data_analysis_details}
    \vspace{-5pt}
\end{table*}

\begin{table*}[!ht]
    \centering
    
    \setlength{\tabcolsep}{3pt}
    \resizebox{1.0\textwidth}{!}{
        \begin{tabular}{l|c|c|c|ccccccc|cc|ccc}
            \toprule
            \multirow{2}{*}{\textbf{Method}} & \multirow{2}{*}{\textbf{Training}} & \multirow{2}{*}{\textbf{Reasoning}} & \multirow{2}{*}{\textbf{Avg.}} & \multicolumn{7}{c|}{\textbf{StackExchange}} & \multicolumn{2}{c|}{\textbf{Coding}} & \multicolumn{3}{c}{\textbf{Theorem-based}} \\
            \cmidrule{5-16}
            & & & & \textbf{Bio.} & \textbf{Earth.} & \textbf{Econ.} & \textbf{Psy.} & \textbf{Rob.} & \textbf{Stack.} & \textbf{Sus.} & \textbf{Leet.} & \textbf{Pony} & \textbf{AoPS} & \textbf{TheoQ.} & \textbf{TheoT.} \\
            \midrule
            \multicolumn{16}{l}{\textbf{{\embed-Qwen3-8B from basic contrastive learning (using InfoNCE loss)}}} \\
            \midrule
            \underline{\textsc{Default}}    & distilled     & w/ reasoning  & 37.1 & 54.4 & 55.4 & 33.8 & 45.2 & 32.0 & 34.3 & 37.3 & 32.3 & 18.7 & 13.3 & 41.2 & 47.6 \\
            \underline{\textsc{Zero-shot}}  & zero-shot     & w/ reasoning  & 32.4 & 46.1 & 51.2 & 31.7 & 40.1 & 28.2 & 31.4 & 32.6 & 33.3 & 4.3  & 9.4  & 38.1 & 42.7 \\
            \underline{\textsc{Non-Reason}} & distilled     & w/o reasoning & 35.0 & 46.4 & 49.6 & 32.2 & 46.9 & 32.4 & 34.2 & 34.4 & 37.2 & 12.6 & 8.3  & 38.6 & 47.1 \\
            \bottomrule
        \end{tabular}
    }
    \caption{Detailed evaluation results (nDCG@10) on \bright benchmark (using original queries) for ablation study of \textbf{relevance annotation} methods.}
    \label{tab:ablation:relevance_annotation_details}
    \vspace{-5pt}
\end{table*}

\begin{table*}[!ht]
    \centering
    \small
    
    \setlength{\tabcolsep}{3pt}
        \begin{tabular}{l|c|ccccccc|cc|ccc}
            \toprule
            \multirow{2}{*}{\textbf{LLM Reasoner}} & \multirow{2}{*}{\textbf{Avg.}} & \multicolumn{7}{c|}{\textbf{StackExchange}} & \multicolumn{2}{c|}{\textbf{Coding}} & \multicolumn{3}{c}{\textbf{Theorem-based}} \\
            \cmidrule{3-14}
            & & \textbf{Bio.} & \textbf{Earth.} & \textbf{Econ.} & \textbf{Psy.} & \textbf{Rob.} & \textbf{Stack.} & \textbf{Sus.} & \textbf{Leet.} & \textbf{Pony} & \textbf{AoPS} & \textbf{TheoQ.} & \textbf{TheoT.} \\
            \midrule
            \multicolumn{14}{l}{\textbf{{\embed-Qwen3-8B from \train (using the self-adaptive RI-InfoNCE loss)}}} \\
            \midrule
            GPT-4.1-mini (default)  & 38.1 & 55.5 & 56.6 & 36.2 & 47.4 & 35.3 & 36.6 & 39.1 & 33.6 & 16.4 & 12.5 & 41.4 & 47.2 \\
            Qwen3-32B               & 37.8 & 55.4 & 56.4 & 35.8 & 48.7 & 33.7 & 37.3 & 37.9 & 33.1 & 16.6 & 12.1 & 42.1 & 45.0 \\
            Qwen3-8B                & 37.5 & 55.3 & 55.1 & 34.2 & 47.5 & 35.0 & 34.5 & 38.4 & 32.6 & 17.7 & 11.8 & 41.4 & 46.1 \\
            Qwen3-4B                & 36.5 & 52.7 & 52.1 & 33.9 & 44.6 & 30.1 & 33.0 & 37.5 & 32.7 & 19.7 & 13.6 & 41.1 & 46.9 \\
            \bottomrule
        \end{tabular}
    \caption{Detailed evaluation results (nDCG@10) on \bright benchmark (using original queries) for ablation study of \textbf{reasoning-intensity} computation methods.}
    \label{tab:ablation:training_details}
    \vspace{-5pt}
\end{table*}

\begin{table*}[!ht]
    \centering
    
    \setlength{\tabcolsep}{3pt}
    \resizebox{1.0\textwidth}{!}{
        \begin{tabular}{l|c|c|ccccccc|cc|ccc}
            \toprule
            \multirow{2}{*}{\textbf{Percentage}} & \multirow{2}{*}{\textbf{Data Size}} & \multirow{2}{*}{\textbf{Avg.}} & \multicolumn{7}{c|}{\textbf{StackExchange}} & \multicolumn{2}{c|}{\textbf{Coding}} & \multicolumn{3}{c}{\textbf{Theorem-based}} \\
            \cmidrule{4-15}
            & & & \textbf{Bio.} & \textbf{Earth.} & \textbf{Econ.} & \textbf{Psy.} & \textbf{Rob.} & \textbf{Stack.} & \textbf{Sus.} & \textbf{Leet.} & \textbf{Pony} & \textbf{AoPS} & \textbf{TheoQ.} & \textbf{TheoT.} \\
            \midrule
            \multicolumn{15}{l}{\textbf{{\embed-Qwen3-8B from basic contrastive learning (using InfoNCE loss)}}} \\
            \midrule
            100.0\%     & 81.6K   & 37.1 & 54.4 & 55.4 & 33.8 & 45.2 & 32.0 & 34.3 & 37.3 & 32.3 & 18.7 & 13.3 & 41.2 & 47.6 \\
            50.0\%      & 40.8K   & 36.6 & 56.2 & 53.0 & 34.4 & 45.6 & 31.9 & 36.4 & 37.6 & 33.8 & 11.4 & 11.9 & 41.7 & 45.7 \\
            25.0\%      & 20.4K   & 36.1 & 54.3 & 54.4 & 33.1 & 44.8 & 30.7 & 35.5 & 35.7 & 34.7 & 8.7  & 12.4 & 41.3 & 48.2 \\
            12.5\%      & 10.2K   & 33.1 & 49.0 & 48.8 & 29.8 & 41.2 & 28.6 & 34.5 & 33.0 & 35.6 & 5.5  & 11.0 & 39.3 & 40.5 \\
            \bottomrule
        \end{tabular}
    }
    \caption{Detailed evaluation results (nDCG@10) on \bright benchmark (using original queries) for ablation study of \textbf{training data size scaling}.}
    \label{tab:ablation:training_data_size_details}
    \vspace{-5pt}
\end{table*}

\end{document}